\RequirePackage{fix-cm}
\documentclass[smallextended]{svjour3}       % onecolumn (second format)
\smartqed  % flush right qed marks, e.g. at end of proof
\usepackage{graphicx}
\usepackage{graphicx}
\usepackage{setspace}
\usepackage{xcolor}
\usepackage[utf8]{inputenc}
\definecolor{red}{cmyk}{0, 0.7808, 0.4429, 0.1412}
\usepackage{float}
\usepackage{siunitx}
\usepackage{bm}
\usepackage{amsmath,amssymb}

\newcommand{\sech}{\mbox{sech}}
\newcommand{\Del}{\Delta}
\newcommand{\del}{\delta}
\newcommand{\al}{\alpha}

\newcommand{\Lam}{\Lambda}
\newcommand{\lam}{\lambda}

\newcommand{\ga}{\gamma}

\newcommand{\bea}{\begin{eqnarray}}
\newcommand{\eea}{\end{eqnarray}}
\newcommand{\bes}{\begin{subequations}}
\newcommand{\ees}{\end{subequations}}
\journalname{NonlinearDyn}
\begin{document}

\title{Coupled Nonlinear Schr\"{o}dinger System: Role of Four-Wave Mixing Effect on Nondegenerate Vector Solitons}

\titlerunning{Nondegenerate solitons in GCNLS system}

\author{R. Ramakrishnan \and 
		M. Kirane \and \\
		S. Stalin \and 
		M. Lakshmanan 
}

\institute{R. Ramakrishnan (Corresponding Author) \at
           Department of Nonlinear Dynamics, 
	       Bharathidasan University, 
           Tiruchirappalli-620 024, 
		   Tamil Nadu, India\\
           \email{ramakrishnan.cnld@gmail.com}
           \and 
           M. Kirane \at
           Department of Mathematics, 
           College of Art and Sciences,
		   Khalifa University of Science and Technology, 
		   Abu Dhabi, United Arab Emirates\\
		   \email{mokhtar.kirane@ku.ac.ae}
	       \and
		   S. Stalin \at
           Department of Mathematics, 
           College of Art and Sciences,
		   Khalifa University of Science and Technology, 
		   Abu Dhabi, United Arab Emirates\\
		   \email{stalin.cnld@gmail.com}
           \and
           M. Lakshmanan \at
           Department of Nonlinear Dynamics, 
	       Bharathidasan University, 
           Tiruchirappalli-620 024, 
		   Tamil Nadu, India\\
           \email{lakshman.cnld@gmail.com} 
}

\date{Received: date / Accepted: date}

\maketitle

\begin{abstract}
In this paper, we investigate the role of four-wave mixing effect on the structure of nondegenerate vector solitons and their collision dynamics.  For this purpose, we consider the generalized coupled nonlinear Schr\"{o}dinger (GCNLS) system which describes the evolution and nonlinear interaction of the two optical modes. The fundamental as well as higher-order nondegenerate vector soliton solutions are derived through the Hirota bilinear method and their forms are rewritten in a compact way using Gram determinants. Very interestingly, we find that the presence of four-wave mixing effect induces a breathing vector soliton state in both the optical modes. Such breather formation is not possible in the fundamental vector bright solitons of the Manakov system. Then, for both strong and weak four-wave mixing effects, we show that  the nondegenerate solitons in the GCNLS system undergo, in general, novel shape changing collisions, in addition to shape preserving collision under suitable choice of wave numbers. Further, we analyze the degenerate soliton collision induced novel shape changing property of nondegenerate vector soliton by deriving the partially nondegenerate two-soliton solution. For completeness, the various collision scenarios related to the pure degenerate bright solitons are indicated. We believe that the results reported in this paper will be useful in nonlinear optics for manipulating light by light through collision.
%\subclass{MSC code1 \and MSC code2 \and more}
\end{abstract}

\section{Introduction}
\label{intro}

In nonlinear optics, some of the most fascinating and intriguing nonlinear phenomena that were observed can be shown to arise due to the nontrivial interactions of light waves \cite{kiv-book,gp}. Among many, the four wave mixing (FWM) is a nonlinear phase sensitive effect in which the interaction of the two copropagating light waves with distinct fundamental frequency components, $\omega_1$ and $\omega_2$, generate new waves \cite{gp}. The frequencies of these new waves (Stokes and anti-Stokes waves) are $\omega_3=\omega_1-\Delta\omega$ and $\omega_4=\omega_2+\Delta\omega$, where $\Delta \omega=\omega_2-\omega_1$. The emergence of the Stokes and anti-Stokes waves mainly depends on the phase matching condition \cite{ansari}. For instance, in Kerr media, the third-order nonlinear susceptibility tensor ($\chi^{(3)}$) results in this parametric process involving four optical waves. Out of these, two pump waves having fundamental frequencies generate anti-Stokes and Stokes side waves, having sum and difference frequencies, and they should obey energy-conservation or phase-matching condition $\omega_1+\omega_2=\omega_3+\omega_4$ \cite{ansari,chen1,chen2}. It is well known that the FWM phenomenon has considerable physical relevance and practical applications particularly in nonlinear optics, especially in supercontinuum generation \cite{alfano,herrmann}, parametric amplification \cite{parametric}, Raman spectroscopy, optical image processing \cite{pepper}, phase conjugate optics \cite{yariv}, etc. This interesting parametric process has been rigorously investigated in the context of spatial and temporal solitons \cite{kiv-book,yulin,nail-cherenkov,husakou1,skryabin}, and in Bose-Einstein condensates (BECs) \cite{wu}.  The dynamics of spatial and temporal solitons with FWM mixing effect is governed by coupled nonlinear Schr\"odinger (CNLS) family of equations with phase dependent nonlinearities. Such CNLS family of equations are, in general, non-integrable in nature. By analyzing these CNLS equations and their corresponding soliton solutions, several interesting results were brought out, including the multicolor solitons \cite{andersen,kiv-multicolor,sammut}. 

On the other hand, considering the physical importance of the FWM effect, completely integrable CNLS equations with phase-dependent nonlinearities have been proposed in different physical contexts. For instance, in nonlinear optics, coherently coupled nonlinear Schr\"odinger equations have been proposed to study the dynamics of two copropagating optical waves in a weakly nonlinear Kerr medium \cite{w1}, electromagnetic wave propagation in gyrotropic nonlinear medium \cite{nail}, matter-wave dynamics in spinor BECs \cite{ieda,pgkeverekidis,keverekidis,kevrekidis1} under special choice of inter and intra-species nonlinear interactions, and propagation of two optical pulses in an isotropic nonlinear Kerr medium \cite{park}. Therefore, understanding the effect of FWM on the vector solitons within the framework of integrable CNLS equations is an important topic in the field of vector solitons with applications in nonlinear optics and BECs. Apart from the latter cases, in Ref. \cite{yang1}, an alternate form of the completely integrable CNLS model with the general form of phase dependent nonlinearity has been proposed to model the propagation and interaction of two optical modes. The form of such  generalized coupled nonlinear Schr\"{o}dinger equations is given by
  \bea
  iq_{j,z}+q_{j,tt}+2Q(q_1,q_2)q_j=0,~q_{j}\equiv q_j(z,t),~j=1,2.  ~~~\label{1}
  %iq_{1,z}+q_{1,tt}+2(a|q_{1}|^{2}+c|q_{2}|^{2}+bq_{1}q_{2}^{*}+b^{*}q_{1}^{*}q_{2})q_{1}=0\\ \nonumber 
  %iq_{2,z}+q_{2,tt}+2(a|q_{1}|^{2}+c|q_{2}|^{2}+bq_{1}q_{2}^{*}+b^{*}q_{1}^{*}q_{2})q_{2}=0 
  \eea               
In the above, $Q(q_1,q_2)=a|q_{1}|^{2}+c|q_{2}|^{2}+bq_{1}q_{2}^{*}+b^{*}q_{1}^{*}q_{2}$, where $q_{j}$'s are the complex light wave envelops, $z$ and $t$ denote the normalized distance and retarded time, respectively. In Eq. (\ref{1}), the real constants $a$ and $c$ describe the self phase modulation (SPM) and cross phase modulation (XPM) effects, respectively, while the complex constant $b$, in the additional phase dependent nonlinearity $(bq_1q_2^*+b^*q_1^*q_2)q_j$, represents the four wave mixing effect. For equal strengths of SPM and XPM effects, that is $a=c$, and $b=0$, the system (\ref{1}) reduces to the Manakov equation \cite{manakov} and the mixed CNLS system \cite{zakharov,mix} arises for $a=-c$, and $b=0$. It was shown that the GCNLS system (\ref{1}) is completely integrable for arbitrary choice of system parameters by providing its Lax pair \cite{yang1} and in Ref. \cite{lu} the Painlev\'e integrability of the system (\ref{1}) was also proved through the Weiss-Tabor-Carnevalle singularity structure algorithm \cite{wtc-algorithm}. Through the Riemann-Hilbert formulation  $N$-bright-bright soliton solutions were obtained and soliton reflection phenomenon was observed therein \cite{yang1}, and also using the Hirota bilinear method $N$-bright-bright and $N$-dark-dark soliton solutions were reported for the system (\ref{1}) \cite{senthil}.

It is interesting to point out that the GCNLS system (\ref{1}) can be mapped to the fundamental vector CNLS models through a simple linear transformation \cite{agalarov},
\begin{equation}
	q_1=\psi_1-b^*\psi_2, ~~q_2=a\psi_2, \label{2}
\end{equation}
or through a general linear transformation \cite{Yuan},
\begin{subequations}
	\begin{eqnarray}
		q_1&=&\alpha_{1}\psi_{1}-(\alpha_{1}^{*}b^{*}+\alpha_{2}^{*}c)\psi_{2}, \label{linear1a}\\ 
		q_2&=&\alpha_{2}\psi_{1}+(\alpha_{1}^{*}a+\alpha_{2}^{*}b)\psi_{2}, \label{linear1b}
	\end{eqnarray}
\end{subequations}
or through even more general linear transformation \cite{Gadzhimuradov},
\begin{subequations}
	\begin{eqnarray}
		&&q_1=\frac{c_1\alpha_jc_2^*}{c_1^*}\psi_1+\alpha_1\psi_2,~c_1=\alpha_1a+\alpha_2b^*,\label{linear2a}\\
		&&q_2=-c_1\alpha_j\psi_1+\alpha_2\psi_2,~c_2=\alpha_1b+\alpha_2c.\label{linear2b}
	\end{eqnarray}
\end{subequations}
In the above, $\psi_j$, $j=1,2$, are the solutions of the fundamental vector CNLS models \cite{agalarov,Yuan,Gadzhimuradov},  and the constants, $\alpha_1$ and $\alpha_2$ are arbitrary complex parameters. By utilizing the above linear connections (2), (3) and (4) one can map the various types of soliton solutions of the fundamental vector CNLS models to the GCNLS system (\ref{1}). For instance, using the transformation (\ref{2}), the bright-bright, dark-dark and quasi-breather-dark soliton solutions were derived for Eq. (\ref{1}). Due to this congruent transformation (\ref{2}), in Ref. \cite{agalarov}, the authors have observed an unconventional dynamics where the density of the first component oscillates in time and space while the second component does not. Then, through transformation (\ref{linear1a})-(\ref{linear1b}), $N$-bright-dark solitons and bound soliton states are studied \cite{Yuan}. By exploiting the transformation (\ref{linear2a})-(\ref{linear2b}), the vector dark solitons with oscillating background have been studied in Ref. \cite{Gadzhimuradov}. In view of the above studies, it is also possible that through the connections (2), (3) and (4) one can also bring out a distinct class of nondegenerate vector soliton solutions, possessing unique behaviors, to the GCNLS system (\ref{1}). Such class of solutions will be reported separately \cite{stalin8} and are essentially different from our soliton solutions derived directly from the GCNLS system (\ref{1}) through the bilinear method, as described below.

Further, it is important to note that in Ref. \cite{manakov} Manakov investigated the two-component solitons in a birefringent fiber or two-mode optical fiber by neglecting the FWM effect. The latter study on vector solitons was based on the completely integrable CNLS equations. The vector bright solitons of such integrable two-CNLS equations without FWM effect undergo a fascinating energy sharing collision through energy redistribution among the modes \cite{radha}. A similar integrable fundamental CNLS model, without FWM terms, was investigated and their various soliton solutions have been extensively studied in Refs. \cite{ablowitz,radha1,boris,boris1,sheppard,kanna,feng,ohta,biondini-2015,prinari-studapplmath,biondini-jmp,biondini-siam}. As we have pointed out earlier in Refs. \cite{ss,ramakrishnan}, in the present GCNLS system (\ref{1}) as well as in the latter mentioned CNLS family of equations \cite{ss1,stalin-review}, the already known energy sharing collision exhibiting solitons are characterized by identical propagation constants. These vector bright solitons are designated as degenerate vector bright solitons. To avoid the degeneracy in the structure of such bright solitons, we choose two distinct wave numbers in the solution construction process. Consideration of the latter choice yields an interesting class of vector bright soliton solutions, which we referred to as the nondegenerate bright solitons \cite{ss,stalin-review}. We note here that in system (\ref{1}) even the vector bright solitons with identical propagation constants in all the modes exhibit energy sharing collisions apart from an interesting soliton reflection-like collision \cite{yang1} where the FWM parameter $b$ plays a crucial role. We wish to point out that the several properties associated with these degenerate vector solitons of the present GCNLS system (\ref{1}) are well understood in the literature. However, to the best of our knowledge the exact analytical forms associated with the fundamental vector bright soliton with two distinct propagation constants as well as the nondegenerate higher-order vector solitons have not been brought out so far in the literature. Also the role of FWM effect on the  propagational and collisional properties of this new class of vector bright solitons have not been explored. The main objective of this paper is to obtain the analytical forms of the nondegenerate vector bright solitons, unveil the role of FWM effect on this special class of vector solitons, and unravel their collision dynamics. We wish to point out that the nondegenerate vector soliton solutions for other integrable CNLS family of systems have also been reported recently by us using the Hirota bilinear method \cite{ss1,stalin-review}. Then, multihump profile structures of this class of nondegenerate soliton solutions  in $N$-CNLS system have been revealed in \cite{stalin5}. We also wish to note that in Ref. \cite{mss} Gadzhimuradov has provided another point of view for understanding the nondegenerate  vector solitons through the linear interference of degenerate vector solitons. This linear interference phenomenon always occurs whenever the wave parameters of degenerate $2N$-soliton solution obeys the so-called interference proposition. One can get the nondegenerate solitons if this interference proposition condition is violated by the wave parameters. Further, we have also shown that the $\mathcal{PT}$-symmetric nonlocal two coupled NLS system also admits both nondegenerate and degenerate soliton solutions \cite{stalin4}. It is interesting to further point out that in the context of BEC using Darboux transformation method, the nondegenerate and degenerate bright and dark solitons have been discussed in Ref. \cite{Qin1,Qin2,Chen1,Chen2,Chen3,Wen}. The nondegenerate soliton solutions and their several properties have been brought out in several coupled systems as they are given as follows: The coupled Fokas-Lenells system \cite{tian}, the two component AB system, \cite{Gao2}, the two-component long-wave short-wave resonance interaction (LSRI) system \cite{stalin-lsri}, the two-component LSRI system of Newell type \cite{jsHe-lsri}, the two coupled mixed derivative nonlinear Schr\"{o}dinger equations \cite{Geng1}, the nonlocal nonlinear Schr\"{o}dinger equation \cite{Geng2}. The effect of inhomogeneity on nondegenerate solitons was studied by considering the variable coefficient Manakov system \cite{Guoli}. 

In order to explore the role of FWM effect on the structural, propagational and collisional properties of the vector nondegenerate solitons in the present paper, we first obtain the exact analytical forms of the fundamental and higher-order nondegenerate vector bright soliton solutions \cite{arxiv-version} by using the well known standard Hirota's bilinear method \cite{hirota-book}. Their general analytical forms are written using the Gram determinants. We find that the presence of phase dependent nonlinearity in the GCNLS system (\ref{1}) induces a novel breathing nondegenerate fundamental soliton state. Then, under strong and weak FWM effects, such breathing nondegenerate solitons undergo a novel shape changing collision and a shape preserving collision, depending on the nature of the  parameters $k_j$, and $l_j$, $j=1,2$.  Furthermore, by restricting these wave numbers appropriately we are able to deduce another class of two-soliton solution, namely partially nondegenerate two-soliton solution, from the completely nondegenerate two-soliton solution. This class of solution is responsible for the coexisting degenerate and nondegenerate solitons. As a result of this coexistence, one is able to study their collision dynamics. By doing so, we identify two types of energy sharing collisions between the degenerate soliton and nondegenerate soliton. In addition to these, we also indicate the various interactions among the two degenerate solitons. To capture these collision scenarios, one has to further impose restriction on the wave numbers.         

The rest of the paper is organized as follows. In Section 2, we present the nondegenerate fundamental and two-soliton solutions through the Hirota bilinear method. Then, in this section, we also point out the existence of partially nondegenerate two-soliton solution and pure degenerate two-soliton solution by imposing restrictions on the wave numbers. The strong and weak FWM effects on the collision properties associated with the nondegenerate solitons are explained in Section 3 with the help of asymptotic analysis. In Section 4, we bring out two types of energy sharing collisions between the degenerate and nondegenerate solitons and indicate the various collision scenarios of the degenerate solitons in Section 5. The results are summarized in Section 6. We present the nondegenerate $N$-soliton solution in Appendix A and the constants that are appearing in the asymptotic analysis of Sections 3.1 and 4.1 are presented in Appendix B and C, respectively.

\section{Nondegenerate soliton solutions}
\label{sec:1}
To derive the nondegenerate soliton solutions, we adopt the well known Hirota bilinear method \cite{hirota-book}, in which the considered coupled nonlinear evolution equation (\ref{1}) should be written in the so-called bilinear form. The bilinear form of Eq. (\ref{1}) can be deduced by introducing the bilinear transformation, namely $q_{j}(z,t)=\frac{g^{(j)}(z,t)}{f(z,t)}$, $j=1,2$, in Eq. (\ref{1}). As a result, the following set of bilinear form is obtained. That is,   
\begin{subequations}
\begin{eqnarray}
&&(iD_{z}+D_{t}^{2})g^{(j)}\cdot f=0,~j=1,2,\label{3.2a}\\
&&D_{t}^{2}f\cdot f=2(ag^{(1)}g^{(1)*}+cg^{(2)}g^{(2)*}+bg^{(1)}g^{(2)*}+b^{*}g^{(1)*}g^{(2)}).\label{3.2b}
\end{eqnarray}
\end{subequations}
In the above, $g^{(j)}(z,t)$'s are complex functions and $f(z,t)$ is a real function, while $D_t$ and $D_z$ are the standard Hirota operators \cite{hirota-book}. Before proceeding further, one has to substitute the series expansions, $g^{(j)}=\epsilon g^{(j)}_1+\epsilon^3 g^{(j)}_3+...$, and $f=1+\epsilon^2 f_2+\epsilon^4 f_4+...$, of the unknown functions $g^{(j)}$ and $f$ in the appropriate places of the above bilinear forms and deduce a system of linear partial differential equations (PDEs) at various orders of $\epsilon$. Solving the resultant set of linear PDEs sucessively one can arrive at either the degenerate or nondegenerate multi-soliton solutions of Eq. (\ref{1}) under appropriate choices of initial seed solutions. 
%%%%%%%%%%%%%%%%%%%%%%%%%%%%%%%%%%%%%%%%%%%%%%%%
\begin{figure*}[ht]
\centering
\includegraphics[width=0.85\linewidth]{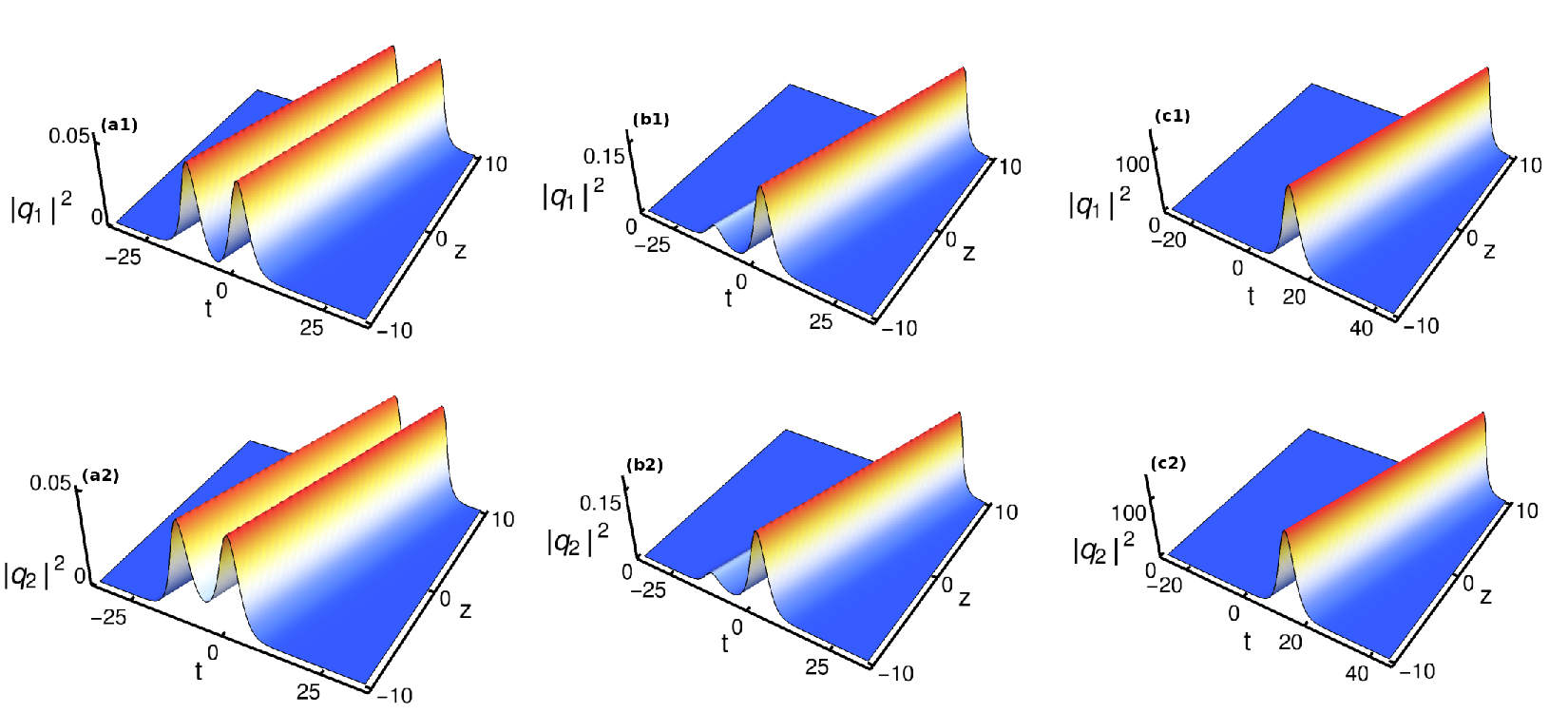}
\caption{The role of FWM effect on the double-hump soliton structure of the nondegenerate one-soliton solution is demonstrated by fixing the parameter values as $k_{1}=0.315+0.5i$, $l_{1}=0.333+0.5i$, $\alpha_{1}^{(1)}=0.5+0.5i$, and $\alpha_{1}^{(2)}=0.45+0.45i$. The strength of FWM for each of the figures. (a1)-(a2): $b=0$, (b1)-(b2): $b=0.5+0.5i$ and (c1)-(c2): $b=1$. }
\label{f1}
\end{figure*}

\subsection{Nondegenerate fundamental vector soliton solution}
To obtain the nondegenerate fundamental soliton solution of Eq. (1), we start with the general form of seed solutions,  $g^{(1)}_{1}=\alpha_{1}^{(1)}e^{\eta_{1}}$,
$g^{(2)}_{1}=\alpha_{1}^{(2)}e^{\xi_{1}}$,  $\eta_{1}=k_{1}t+ik_{1}^{2}z$ and $\xi_{1}=l_{1}t+il_{1}^{2}z$, as the starting solutions to the lowest order linear PDEs, $ig_{1z}^{(j)}+g_{1tt}^{(j)}=0$, $j=1,2$. Here $\alpha_1^{(1)}$, $\alpha_1^{(2)}$, $k_1$ and $l_1$  are arbitrary complex constants and in general $k_1\neq l_1$.  We remark here that the previously known class of fundamental vector soliton solution of the GCNLS system (\ref{1}) can be obtained by considering the limited form of the seed solutions, $g^{(1)}_{1}=\alpha_{1}^{(1)}e^{\eta_{1}}$,
$g^{(2)}_{1}=\alpha_{1}^{(2)}e^{\eta_{1}}$, $\eta_{1}=k_{1}t+ik_{1}^{2}z$, which can be easily deduced from the above general choice with $k_1=l_1$ \cite{senthil}. Then, by following the standard procedure of the Hirota method we arrive at the nondegenerate fundamental bright soliton solution of the system (\ref{1}) as
%\frac{g_{1}^{(1)}+g_{3}^{(1)}}{1+f_{2}+f_{4}}, \frac{g_{1}^{(2)}+g_{3}^{(2)}}{1+f_{2}+f_{4}}
\bes
\begin{eqnarray}
&&q_1=\frac{1}{D}\big(\alpha_{1}^{(1)}e^{\eta_{1}}+e^{\eta_{1}+\eta_{1}^{*}+\xi_{1}+\Delta_{1}^{(1)}}+e^{\eta_{1}+\xi_{1}+\xi_{1}^{*}+\Delta_{2}^{(1)}}\big),~~~~~\label{3.3a}\\	
&&q_2=\frac{1}{D}\big(\alpha_{1}^{(2)}e^{\xi_{1}}+e^{\eta_{1}+\eta_{1}^{*}+\xi_{1}+\Delta_{1}^{(2)}}+e^{\eta_{1}+\xi_{1}+\xi_{1}^{*}+\Delta_{2}^{(2)}}\big),~~~~\label{3.3b}\\
%&&q_{1}=\big(\alpha_{1}^{(1)}e^{\eta_{1}}+e^{\eta_{1}+\eta_{1}^{*}+\xi_{1}+\Delta_{1}^{(1)}}\nonumber\\&&\hspace{1.0cm}+e^{\eta_{1}+\xi_{1}+\xi_{1}^{*}+\Delta_{2}^{(1)}}\big)/D,\label{3.3a}\\
%&&q_{2}=\big(\alpha_{1}^{(2)}e^{\xi_{1}}+e^{\eta_{1}+\eta_{1}^{*}+\xi_{1}+\Delta_{1}^{(2)}}\nonumber\\&&\hspace{1.0cm}+e^{\eta_{1}+\xi_{1}+\xi_{1}^{*}+\Delta_{2}^{(2)}}\big)/D,\label{3.3b}\\
&&D=1+e^{\eta_{1}+\eta_{1}^{*}+\delta_{1}}+e^{\eta_{1}^{*}+\xi_{1}+\delta_{2}}+e^{\eta_{1}+\xi_{1}^{*}+\delta_{2}^*}+e^{\xi_{1}+\xi_{1}^{*}+\delta_{3}}+e^{\eta_{1}+\eta_{1}^{*}+\xi_{1}+\xi_{1}^{*}+\delta_{4}}\nonumber.
\end{eqnarray}\ees
Here, 
$e^{\Delta_{1}^{(1)}}=\frac{b^{*}(k_{1}-l_{1})|\alpha_{1}^{(1)}|^{2}\alpha_{1}^{(2)}}{(k_{1}+k_{1}^{*})(k_{1}^{*}+l_{1})^{2}}$, 
$e^{\Delta_{2}^{(1)}}=\frac{c(k_{1}-l_{1})\alpha_{1}^{(1)}|\alpha_{1}^{(2)}|^{2}}{(k_{1}+l_{1}^{*})(l_{1}+l_{1}^{*})^{2}}$,
$e^{\delta_{1}}=\frac{a|\alpha_{1}^{(1)}|^{2}}{(k_{1}+k_{1}^{*})^{2}}$, 
$e^{\Delta_{1}^{(2)}}=-\frac{a(k_{1}-l_{1})|\alpha_{1}^{(1)}|^{2}\alpha_{1}^{(2)}}{(l_{1}+k_{1}^{*})(k_{1}+k_{1}^{*})^{2}}$, 
$e^{\Delta_{2}^{(2)}}=-\frac{b(k_{1}-l_{1})\alpha_{1}^{(1)}|\alpha_{1}^{(2)}|^{2}}{(l_{1}+l_{1}^{*})(k_{1}+l_{1}^{*})^{2}}$, 
$e^{\delta_{2}}=\frac{b^*\alpha_{1}^{(1)*}\alpha_{1}^{(2)}}{(k_{1}^*+l_{1})^{2}}$, 
$e^{\delta_{3}}=\frac{c|\alpha_{1}^{(2)}|^{2}}{(l_{1}+l_{1}^{*})^{2}}$,
$e^{\delta_{4}}=\frac{|k_{1}-l_{1}|^{2}|\alpha_{1}^{(1)}|^{2}|\alpha_{1}^{(2)}|^{2}\big[ac|k_{1}+l_{1}^{*}|^{2}-|b|^{2}(k_{1}+k_{1}^{*})(l_{1}+l_{1}^{*})\big]}{(k_{1}+k_{1}^{*})^{2}|k_{1}+l_{1}^*|^{4}(l_{1}+l_{1}^{*})^{2}}$. The nature of the above solution is described by four arbitrary complex parameters, $k_1$, $l_1$, $\alpha_1^{(j)}$, $j=1,2$, and three system parameters $a$, $c$ and $b$. Further, in order that the solution (\ref{3.3a})-(\ref{3.3a}) is nonsingular in nature, we require  the denominator terms, $e^{\del_j}$, $j=1,2,3,4$, occurring in the expression for $D$ in the solution (\ref{3.3a})-(\ref{3.3b}) should be positive definite. The latter is true if the strengths of SPM and XPM are positive ($a,c>0$) and the term  $\big(ac|k_{1}+l_{1}^{*}|^{2}-|b|^{2}(k_{1}+k_{1}^{*})(l_{1}+l_{1}^{*})\big)$ is greater than zero.
 
For $b=0$, the solution (\ref{3.3a})-(\ref{3.3b}) exactly coincides with the nondegenerate fundamental bright soliton solution of the Manakov system \cite{ss} and mixed 2-CNLS system \cite{stalin-review} by further fixing  $a=c=1$ and $a=-c=1$, respectively, in it. 
The previously reported three-parameter vector soliton solution of the GCNLS system (\ref{1}) \cite{senthil} arises as a special case when we impose $k_1=l_1$ in the above four-parameter family of solution (\ref{3.3a})-(\ref{3.3b}). As a result, the explicit form of three-parameter bright soliton solution turns out to be   
$q_{j}=\frac{\alpha_{1}^{(j)}e^{\eta_1}}{1+e^{\eta_1+\eta_1^*+R}}\equiv k_{1R} \hat{A_{j}}e^{i\eta_{1I}}\sech(\eta_{1R}+\frac{R}{2})$, $j=1,2$, where $\eta_1=k_1t+ik_1^2z=\eta_{1R}+i\eta_{1I}=[k_{1R}(t-2k_{1I}z)]+i[k_{1I}t+(k_{1R}^{2}-k_{1I}^{2})z]$. Here, the polarization vector $A$ is equal to $\begin{pmatrix}
	\hat{A}_1, &\hat{A}_2
\end{pmatrix}^{T}$, where $\hat{A_{j}} = \alpha_{1}^{(j)}/[a|\alpha_{1}^{(1)}|^2+c|\alpha_{1}^{(2)}|^2+b\alpha_{1}^{(1)}\alpha_{1}^{(2)*}+b^*\alpha_{1}^{(1)*}\alpha_{1}^{(2)}]^{\frac{1}{2}}$, $j=1,2$,  $e^{R}=\frac{(a|\alpha_{1}^{(1)}|^2+c|\alpha_{1}^{(2)}|^2+b\alpha_{1}^{(1)}\alpha_{1}^{(2)*}+b^*\alpha_{1}^{(1)*}\alpha_{1}^{(2)})}{(k_{1}+k_{1}^*)^2}$, the amplitude of the two modes are $k_{1R}\hat{A_{j}}$, the velocity of the degenerate soliton is $2k_{1I}$ and the central position of the soliton is identified as  $\frac{R}{2k_{1R}}=\frac{1}{k_{1R}}\log\frac{(a|\alpha_{1}^{(1)}|^2+c|\alpha_{1}^{(2)}|^2+b\alpha_{1}^{(1)}\alpha_{1}^{(2)*}+b^*\alpha_{1}^{(1)*}\alpha_{1}^{(2)})^{\frac{1}{2}}}{(k_{1}+k_{1}^*)}$. In the present GCNLS system (\ref{1}), the polarization vector of the above degenerate soliton solution, $A\equiv \begin{pmatrix}
	\hat{A}_1, &\hat{A}_2
\end{pmatrix}^{T}$, is said to be a unit polarization vector as it obeys the required relation $A^\dagger BA=1$, $B=\begin{pmatrix}
a &b^*\\ b&c
\end{pmatrix}=B^{\dagger}$ \cite{yang1}. We note that the above degenerate bright soliton solution always admits a single-hump `$\sech$' soliton profile.\\
To bring out the special properties associated with the solution (\ref{3.3a})-(\ref{3.3b}) further, we rewrite it as follows: 
%\begin{widetext}
{\footnotesize
\bes
\bea
q_1&=&\frac{2k_{1R}}{D_1}\bigg(c_{11}e^{i\eta_{1I}}\cosh(\xi_{1R}+\phi_1)+c_{21}e^{i\xi_{1I}}[\cosh(\eta_{1R}+\phi_2-\phi_1+c_2)\nonumber\\
&+&\sinh(\eta_{1R}+\phi_2-\phi_1+c_2)]\bigg),\label{3.4a}\\
q_2&=&\frac{2l_{1R}}{D_1}\bigg(c_{12}e^{i\xi_{1I}}\cosh(\eta_{1R}+\phi_2)+c_{22}e^{i\eta_{1I}}[\cosh(\xi_{1R}-(\phi_2-\phi_1)+c_2)\nonumber\\
&+&\sinh(\xi_{1R}-(\phi_2-\phi_1)+c_2)]\bigg),\label{3.4b}\\
D_1&=&\Lambda_1\cosh(\eta_{1R}+\xi_{1R}+\phi_2+\phi_1+c_1)
+\cosh(\eta_{1R}-\xi_{1R}+\phi_2-\phi_1+c_2)\nonumber\\
&+&\Lambda_2[\cosh \phi_3\cos(\eta_{1I}-\xi_{1I})+i\sinh\phi_3\sin(\eta_{1I}-\xi_{1I})]. \nonumber
\eea
\ees
%\end{widetext}
}
Here, $\eta_{1R}=k_{1R}(t-2k_{1I}z)$, 
$\xi_{1R}=l_{1R}(t-2l_{1I}z)$, 
$\eta_{1I}=k_{1I}t+(k_{1R}^2-k_{1I}^2)z$,  
$\xi_{1I}=l_{1I}t+(l_{1R}^2-l_{1I}^2)z$,  
$\phi_1=\frac{1}{2}\log\frac{c(k_1-l_1)|\alpha_1^{(2)}|^2}{(k_1+l_1^*)(l_1+l_1^*)^2}$, 
$\phi_2=\frac{1}{2}\log\frac{a(l_1-k_1)|\alpha_1^{(1)}|^2}{(k_1^*+l_1)(k_1+k_1^*)^2}$, 
$\phi_3=\frac{1}{2}\log\frac{b\alpha_1^{(1)}\alpha_1^{(2)*}(k_1^*+l_1)^2}{b^*\alpha_1^{(1)*}\alpha_1^{(2)}(k_1+l_1^*)^2}$, 
$c_{11}=\bigg(\frac{\alpha_1^{(1)}(k_1-l_1)}{a\alpha_1^{(1)*}(k_1+l_1^*)}\bigg)^{1/2}$, 
$c_{21}=\frac{1}{2}\bigg(\frac{b^*\alpha_1^{(2)}(k_1-l_1)}{a(k_1^*+l_1)^2}\bigg)$, 
$c_{12}=\bigg(\frac{\alpha_1^{(2)}(l_1-k_1)}{c\alpha_1^{(2)*}(k_1^*+l_1)}\bigg)^{1/2}$, 
$c_{22}=\frac{1}{2}\bigg(\frac{b\alpha_1^{(1)}(l_1-k_1)}{c(k_1+l_1^*)^2}\bigg)$,
$c_2=\frac{1}{2}\log\frac{(k_1-l_1)(k_1^*+l_1)}{(l_1-k_1)(k_1+l_1^*)}$,
$c_1=\frac{1}{2}\log\frac{(k_1^*-l_1^*)[ac|k_1+l_1^*|^2-|b|^2(k_1+k_1^*)(l_1+l_1^*)]}{ac(l_1-k_1)|k_1+l_1^*|^2}$, 
$\Lambda_1=\frac{|k_1-l_1|[ac|k_1+l_1^*|^2-|b|^2(k_1+k_1^*)(l_1+l_1^*)]^{1/2}}{(ac)^{1/2}|k_1+l_1^*|^2}$, and
$\Lambda_2=\frac{|b|(k_1+k_1^*)(l_1+l_1^*)}{(ac)^{1/2}|k_1+l_1^*|^2}$. The presence of additional wave number $k_1$ or $l_1$ provides an extra degree of freedom to the motion as well as to the structure of the soliton in the two modes $q_1$ and $q_2$. For instance, the following two possibilities are always allowed. The solitons in the two modes can propagate with either equal velocities:  $v_1=v_2$, where $v_1=2k_{1I}$, $v_2=2l_{1I}$ or with unequal velocities: $v_1\neq v_2$. As we describe below, these two choices reveal the new geometrical structures related to the solution (\ref{3.3a})-(\ref{3.3b}) of the GCNLS system (\ref{1}).  We wish to note that the formation of nondegenerate one soliton solution (\ref{3.3a})-(\ref{3.3b}) can also be interpreted through the linear interference of degenerate two soliton solution of the GCNLS system (1), as it is explained in Ref. \cite{mss} in the case of Manakov system.
%%%%%%%%%%%%%%%%%%%%%%%%%%%%%%%
\subsubsection{Role of FWM effect on one-soliton solution}
The nondegenerate fundamental soliton solution (\ref{3.3a})-(\ref{3.3b}) with $v_1=v_2$ admits double-hump profile when the FWM effect is zero. Such profiles are displayed in Figs. \ref{1}(a1) and \ref{1}(a2) for $b=0$ and $a=c=1$. However, the symmetric nature of such intensity profiles disappears and  asymmetric double-hump profiles emerge in both the modes $q_1$ and $q_2$ when we incorporate the FWM effect ($b\neq 0$) along with the assignment that the real part of $k_1$ is slightly greater than the real part of $l_1$ $(k_{1R}>l_{1R})$. Such a profile transition is displayed in Figs. \ref{1}(b1) and \ref{1}(b2). On further increasing the value of $b$, we find that the first-hump is completely suppressed in both the modes and the second-hump only persists throughout the evolution  with an enhancement in amplitude or intensity, which is illustrated in Figs. \ref{1}(c1) and \ref{1}(c2).

\begin{figure}[t]
	\centering
	\includegraphics[width=1\linewidth]{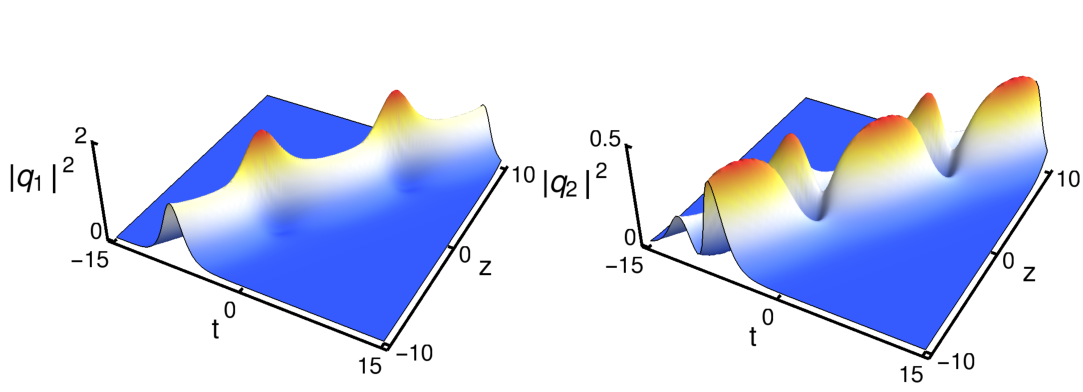}
	\caption{Breather formation with strong FWM effect is demonstrated by fixing the parameter values as $a=c=1$, $b=0.5+0.5i$, $k_{1}=1+0.5i$, $l_{1}=0.5+0.5i$,  $\alpha_{1}^{(1)}=0.65$, and $\alpha_{1}^{(2)}=1+i$. }
	\label{f2}
\end{figure}

\begin{figure}[t]
	\centering
	\includegraphics[width=1\linewidth]{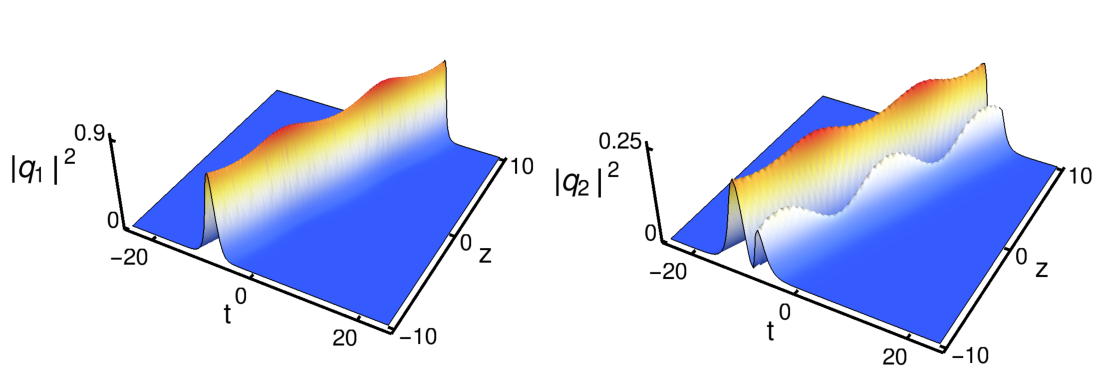}
	\caption{Breathing state is demonstrated for the low strength of FWM effect. The parameter values are the same as  in Fig. \ref{f2} except $b=0.15+0.15i$. }
	\label{f3}
\end{figure}

\begin{figure}[t]
	\centering
	\includegraphics[width=1\linewidth]{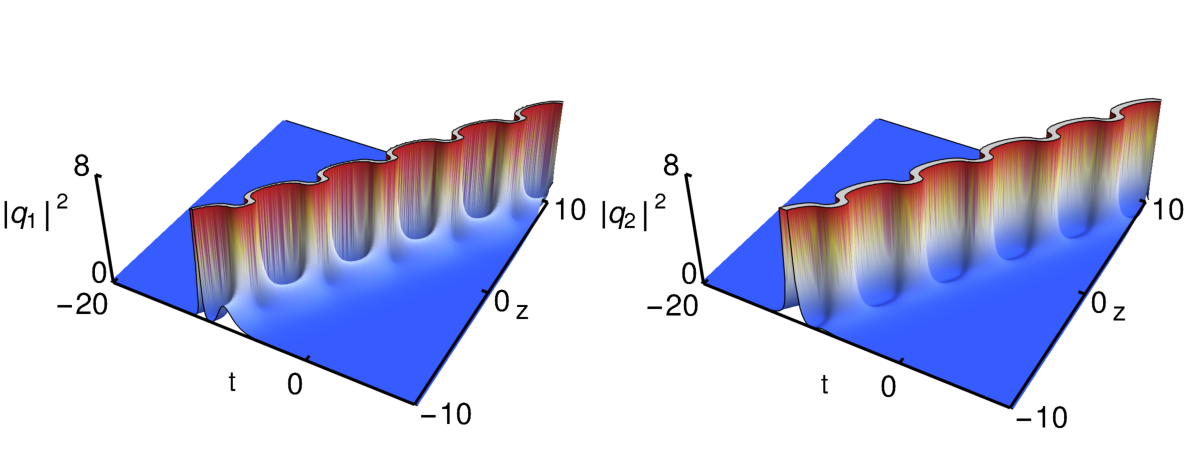}
	\caption{Singular breathing state is demonstrated for the strong FWM effect. The parameter values are  $b=0.5+0.5i$, $k_{1}=1.3+0.5i$, $l_{1}=-0.5+0.5i$,  $\alpha_{1}^{(1)}=0.65$, and $\alpha_{1}^{(2)}=i$. }
	\label{f4}
\end{figure}

Interestingly, we also find that the presence of FWM parameter generates a breathing state in the structure of the nondegenerate fundamental soliton of the GCNLS system (\ref{1}). It can be identified from the expressions (\ref{3.4a})-(\ref{3.4b}) with $v_1=v_2$, where periodic functions explicitly appear because of the complex nature of the FWM parameter $b$. For $b=0$, periodic functions would disappear from Eqs. (\ref{3.4a}) and(\ref{3.4b}) and subsequently the breathing behavior will be absent as in the cases of the Manakov \cite{ss,manakov} and the mixed CNLS system \cite{ss1,mix}. Such novel breathing state in the present GCNLS system is depicted in Figs. \ref{f2} and \ref{f3}, where the oscillations occur along the propagation direction $z$ only. From these figures, we observe that the strong breathing nature appears when the FWM effect is high enough (see Fig. \ref{f2}) along with a parameteric condition $k_{1R}>>l_{1R}$, in which the value of $k_{1R}$ should be considerably larger than $l_{1R}$ (or vice versa). On the other hand, for a weak strength of the FWM effect, the small oscillations appear in the intensity peaks only (see Fig. \ref{f3}). The period of oscillation is calculated as 
\begin{equation}
T=\frac{2\pi}{\omega}=\frac{2\pi}{(k_{1R}^2-l_{1R}^2)}\label{3.5}.
\end{equation}    
The above expression shows that the period of oscillations is mainly dependent on the real parts of the wave numbers $k_1$ and $l_1$ in addition to the FWM nonlinearity ($b$). This type of special property has not been observed in the degenerate counterparts, where the real part of a single wave number $k_1$ describes only the amplitude of the degenerate vector bright soliton of Eq. (\ref{1}) accompanying with a polarization vector. For completeness, in Fig. \ref{f4}, we also demonstrate the breathing singular soliton state by considering the mixed type nonlinearity $a=1$, $c=-1$. As singular solitons are nonphysical, this choice of parameters may be avoided when considering nondegenerate solitons in applications. Here, we note that, in Ref. \cite{priya}, it has been observed that a single-humped bright soliton on the constant wave background gets converted into a breather form while tuning the value of $b$. However, as we pointed out above, in our present case, the double-humped non-degenerate soliton starts to breath when we tune the value of $b$ as well as the real parts of distinct wave numbers $k_1$ and $l_1$.
%%%%%%%%%%%%%%%%%%%%%%%%%%%%%%%%%%%%%%%%%%%%%%%%%%%%%%%%%%%
%%%%%%%%%%%%%%%%%%%%%%%%%%%%%%%%%%%%%%%%%%%%%%%%%%%%%%%%
%%%%%%%%%%%%%%%%%%%%%%%%%%%%%%%%%%%%%%%%%%%%%%%%%%%%
%%%%%%%%%%%%%%%%%%%%%%%%%%%%%%%%%%%%%%%%%%%%%%%
\begin{figure}[t]
	\centering
	\includegraphics[width=1\linewidth]{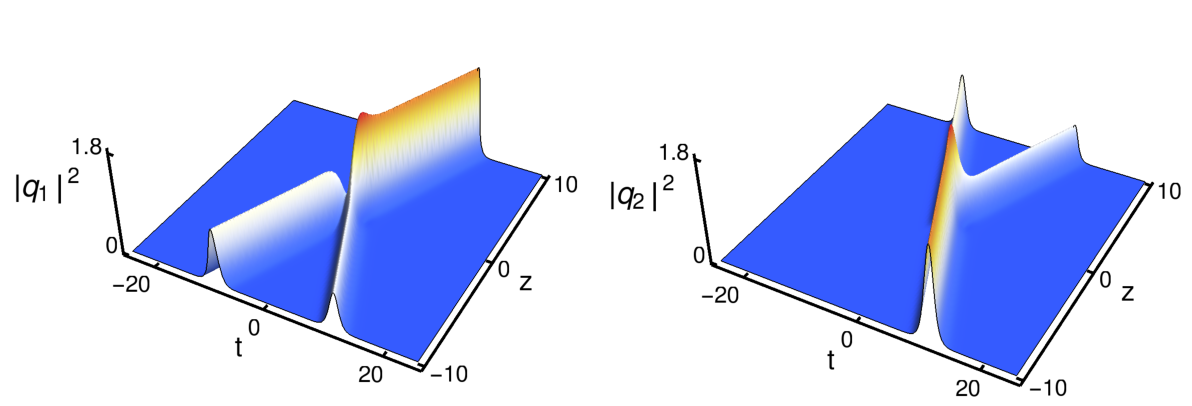}
	\caption{Nondegenerate fundamental soliton with unequal-velocities by fixing the parameter values as $a=c=1$, $b=0.5+0.5i$, $k_{1}=1+0.5i$, $l_{1}=1-0.5i$,  $\alpha_{1}^{(1)}=1$, and $\alpha_{1}^{(2)}=1+i$. }
	\label{f5}
\end{figure}
%%%%%%%%%%%%%%%%%%%%%%%%%%%%%%%%%%%%%%%%%%%%%%%
\begin{figure}[t]
	\centering
	\includegraphics[width=1\linewidth]{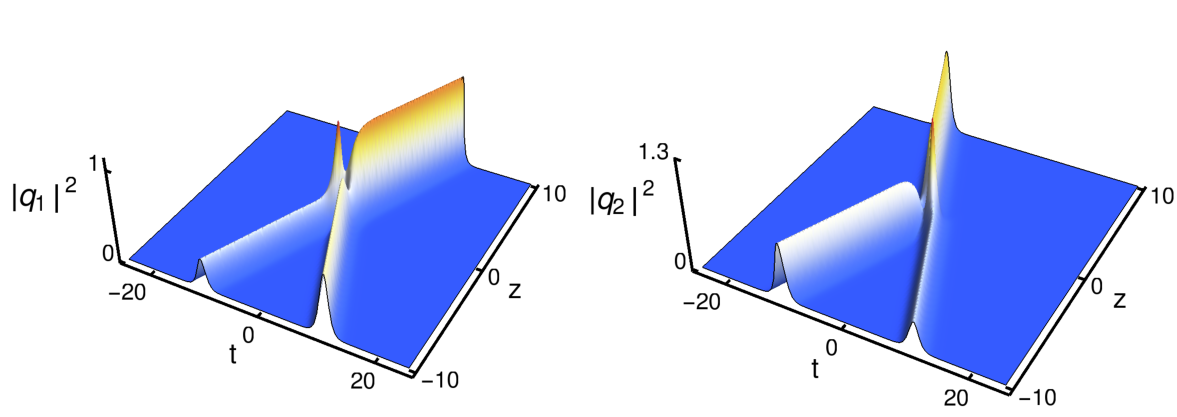}
	\caption{Nondegenerate fundamental soliton with unequal-velocities by fixing the parameter values as $a=c=1$, $b=0.5+0.5i$, $k_{1}=1+0.5i$, $l_{1}=-1-0.5i$,  $\alpha_{1}^{(1)}=1$, and $\alpha_{1}^{(2)}=1+i$. }
	\label{f6}
\end{figure}
%%%%%%%%%%%%%%%%%%%%%%%%%%%%%%%%%%%%%%%%%%%%%%%

Next, we consider the solution (\ref{3.3a})-(\ref{3.3b}) with unequal velocities: $v_1\neq v_2$. In this situation, it admits two types of interesting patterns as we have illustrated in Figs. \ref{f5}  and \ref{f6}. In these figures, two distinct single-hump profiles at different position start to interact at $z=0$. As a result, these interaction patterns appear due to the exchange of intensities among the modes. This kind of switching of intensities among the wave guides could be relevant to optical switching  applications. Further, we wish to state that the numerical stability of nondegenerate soliton solution (6a)-(6b) is confirmed by adding $10\%$, and $20\%$ of white noise, as weak and strong perturbations, to the initial solution as  $q_j(0,t)=[1+A\zeta(t)]q_{j,0}(t)$, $j=1,2$, where $q_{j,0}(t)$ is the solution of the system (1) at $z=0$,  $A$ is the amplitude of the white noise and $\zeta(t)$ is the noise function,   of the GCNLS system (1). From this analysis, we have observed that the profile of non-degenerate soliton does not get distorted and it survives against both strengths of perturbation. Thus, it ensures the stability of non-degenerate soliton solution (6a)-(6b).
       
%%%%%%%%%%%%%%%%%%%%%%%%%%%%%%%%%%%%%%%%%%%%%%%%%
\subsection{Completely/partially nondegenerate two-soliton solution}
Depending on the choices of the seed solutions consideration along with the following conditions on the wave numbers, namely (i) $k_1\neq l_1$ , $k_2\neq l_2$, (ii) $k_1= l_1$ and $k_2\neq l_2$ (or $k_1\neq l_1$ and $k_2=l_2$), and (iii) $k_1= l_1$ and $k_2=l_2$, the GCNLS system (\ref{1}) admits three-types of two-soliton solutions, namely (i) completely nondegenerate two-soliton solution, (ii)  partially nondegenerate two-soliton solution, and (iii) completely degenerate two-soliton solution, respectively. For instance, the two-soliton solution, with  the complete nondegeneracy property, is obtained as a result of finding the unknown functions in the truncated series expansions, $g^{(j)}=\epsilon g_1^{(j)}+\epsilon^3 g_3^{(j)}+\epsilon^5 g_5^{(j)}+\epsilon^7 g_7^{(j)}$, $j=1,2$, and $f=1+\epsilon^2 f_2+\epsilon^4 f_4+\epsilon^6 f_6+\epsilon^8 f_8$. To get the explicit forms of the unknown functions that are present in the latter series expansions, we assume the initial solutions as 
\begin{eqnarray}
&&g_1^{(1)}=\al_1^{(1)}e^{\eta_1}+\al_2^{(1)}e^{\eta_2} ~ \text{and}~ g_1^{(2)}=\al_1^{(2)}e^{\xi_1}+\al_2^{(2)}e^{\xi_2},~~\label{3.6}\\ &&\eta_{j}=k_{j}t+ik_{j}^{2}z,~ \xi_{j}=l_{j}t+il_{j}^{2}z,~ j=1,2.\nonumber
\end{eqnarray}
Here, the wave numbers $k_j$ and $l_j$ and the constants $\alpha_1^{(j)}$ and $\alpha_2^{(j)}$, $j=1,2$, are in general complex. We find that the other unknown functions, $g_9^{(j)}$, $g_{11}^{(j)}$, $j=1,2$, $f_{10}$,  $f_{12}$ and etc. exactly vanish. The remaining non-vanishing functions constitute the nondegenerate two-soliton solution, which is rewritten using the Gram determinants in the following way:
\begin{eqnarray}
\hspace{-0.5cm}g^{(s)}&=&
\begin{vmatrix}
A & I & \phi \\ 
-I &  B & {\bf 0}^T\\
{\bf 0} & C_s & 0
\end{vmatrix},~f=\begin{vmatrix}
A & I \\ 
-I &  B 
\end{vmatrix},~s=1,2,\label{3.7}
\end{eqnarray} where the other elements in the above determinants are defined below:  
%%%%%%%%%%%%%%%%%%%%%%%%%%%%%%%%%

\begin{eqnarray}
&&A =
\begin{pmatrix}
\frac{e^{\eta_1+\eta_1^*}}{(k_1+k_1^*)} & \frac{e^{\eta_1+\eta_2^*}}{(k_1+k_2^*)} & \frac{e^{\eta_1+\xi_1^*}}{(k_1+l_1^*)} & \frac{e^{\eta_1+\xi_2^*}}{(k_1+l_2^*)} \\ 
\frac{e^{\eta_2+\eta_1^*}}{(k_2+k_1^*)} & \frac{e^{\eta_2+\eta_2^*}}{(k_2+k_2^*)} & \frac{e^{\eta_2+\xi_1^*}}{(k_2+l_1^*)} & \frac{e^{\eta_2+\xi_2^*}}{(k_2+l_2^*)} \\ 
\frac{e^{\xi_1+\eta_1^*}}{(l_1+k_1^*)} & \frac{e^{\xi_1+\eta_2^*}}{(l_1+k_2^*)} & \frac{e^{\xi_1+\xi_1^*}}{(l_1+l_1^*)} & \frac{e^{\xi_1+\xi_2^*}}{(l_1+l_2^*)} \\ 
\frac{e^{\xi_2+\eta_1^*}}{(l_2+k_1^*)} & \frac{e^{\xi_2+\eta_2^*}}{(l_2+k_2^*)} & \frac{e^{\xi_2+\xi_1^*}}{(l_2+l_1^*)} & \frac{e^{\xi_2+\xi_2^*}}{(l_2+l_2^*)} 
\end{pmatrix},~\nonumber\\
&&B=\begin{pmatrix}
\frac{a\alpha_1^{(1)}\alpha_1^{(1)*}}{(k_1+k_1^*)} & \frac{a\alpha_2^{(1)}\alpha_1^{(1)*}}{(k_2+k_1^*)} & \frac{b^*\alpha_1^{(2)}\alpha_1^{(1)*}}{(l_1+k_1^*)}&\frac{b^*\alpha_2^{(2)}\alpha_1^{(1)*}}{(l_2+k_1^*)}\\ 
\frac{a\alpha_1^{(1)}\alpha_2^{(1)*}}{(k_1+k_2^*)} & \frac{a\alpha_2^{(1)}\alpha_2^{(1)*}}{(k_2+k_2^*)} & \frac{b^*\alpha_1^{(2)}\alpha_2^{(1)*}}{(l_1+k_2^*)}&\frac{b^*\alpha_2^{(2)}\alpha_2^{(1)*}}{(l_2+k_2^*)} \\ 
\frac{b\alpha_1^{(1)}\alpha_1^{(2)*}}{(k_1+l_1^*)} & \frac{b\alpha_2^{(1)}\alpha_1^{(2)*}}{(k_2+l_1^*)} & \frac{c\alpha_1^{(2)}\alpha_1^{(2)*}}{(l_1+l_1^*)}&\frac{c\alpha_2^{(2)}\alpha_1^{(2)*}}{(l_2+l_1^*)}\\
\frac{b\alpha_1^{(1)}\alpha_2^{(2)*}}{(k_1+l_2^*)} & \frac{b\alpha_2^{(1)}\alpha_2^{(2)*}}{(k_2+l_2^*)} & \frac{c\alpha_1^{(2)}\alpha_2^{(2)*}}{(l_1+l_2^*)}&\frac{c\alpha_2^{(2)}\alpha_2^{(2)*}}{(l_2+l_2^*)}
\end{pmatrix},\nonumber\\
&&\phi=\begin{pmatrix}
e^{\eta_1} & e^{\eta_2}  & e^{\xi_1}  & e^{\xi_2} 
\end{pmatrix}^T,~C_1=-\begin{pmatrix}
\alpha_1^{(1)} & \alpha_2^{(1)} & 0 & 0 
\end{pmatrix},\nonumber\\&&C_2=-\begin{pmatrix}
 0 & 0 &\alpha_1^{(2)} & \alpha_2^{(2)}  \end{pmatrix},~{\bf 0} =\begin{pmatrix}
 0 & 0 & 0 &0 & \end{pmatrix},
\end{eqnarray}

and $I$ is a $(4\times 4)$ identity matrix. The above solution consists of eight arbitrary complex parameters $k_{j}$, $l_{j}$, $\alpha_{1}^{(j)}$ and $\alpha_{2}^{(j)}$, $j=1,2$. The profile shapes of the nondegenerate solitons and their various novel collision scenarios are governed by these eight nontrivial soliton parameters and the three system parameters $a$, $c$ and $b$. By generalizing the procedure given above, the nondegenerate $N$-soliton solution of the GCNLS system can also be obtained and its explicit form is given in \ref{A}. For this purpose one has to consider the more general forms of seed solutions as \bea
\hspace{-0.5cm}&&g_1^{(1)}=\sum_{j=1}^{N}\alpha_{j}^{(1)}e^{\eta_j},~~~~~ g_1^{(2)}=\sum_{j=1}^{N}\alpha_{j}^{(2)}e^{\xi_j},\label{3.8}\\ \hspace{-0.5cm}&&\eta_{j}=k_{j}x+ik_{j}^{2}t,~~~~~\xi_{j}=l_{j}x+il_{j}^{2}t,~j=1,2,...,N.\nonumber
\eea 
Therefore, the resultant $N$-soliton solution (\ref{3.A1}) contains $4N$-complex parameters, $k_j$, $l_j$, $\alpha_1^{(j)}$, and $\alpha_2^{(j)}$, $j=1,2,...,N$. 

In addition, we wish to point out that the GCNLS system (\ref{1}) also admits another class of two-soliton solution containing both degenerate and nondegenerate vector solitons simultaneously. This additional possibility always exists in the newly derived two-soliton solution (\ref{3.7}). Such a possibility arises by restricting the sets of wave numbers as $k_1=l_1$ and $k_2\neq l_2$ or $k_1\neq l_1$ and $k_2=l_2$ in Eq. (\ref{3.7}). Here, we have considered the former choice. By doing so, the seed solutions (\ref{3.6}) get reduced as \begin{eqnarray}
&&g_1^{(1)}=\al_1^{(1)}e^{\eta_1}+\al_2^{(1)}e^{\eta_2}, ~ g_1^{(2)}=\al_1^{(2)}e^{\eta_1}+\al_2^{(2)}e^{\xi_2},~~\label{3.9}\\ &&\eta_{j}=k_{j}t+ik_{j}^{2}z,~ \text{and}~ \xi_{2}=l_{2}t+il_{2}^{2}z,~ j=1,2.\nonumber
\end{eqnarray} 
With the above choice of initial solutions one can also derive the partial nondegenerate two-soliton solution through the Hirota bilinear method. We obtain the following form of the partial nondegenerate two-soliton solution as a final product. However, the resultant form is same as the one given in Eq. (\ref{3.7}) except for the changes that occur in the elements of the matrices $A$, $B$ and $\phi$ as given below:   
%%%%%%%%%%%%%%%%%%%%%%%%%%%%%%%%%

	\begin{eqnarray}
	&&A =
	\begin{pmatrix}
	\frac{e^{\eta_1+\eta_1^*}}{(k_1+k_1^*)} & \frac{e^{\eta_1+\eta_2^*}}{(k_1+k_2^*)} & \frac{e^{\eta_1+\eta_1^*}}{(k_1+k_1^*)} & \frac{e^{\eta_1+\xi_2^*}}{(k_1+l_2^*)} \\ 
	\frac{e^{\eta_2+\eta_1^*}}{(k_2+k_1^*)} & \frac{e^{\eta_2+\eta_2^*}}{(k_2+k_2^*)} & \frac{e^{\eta_2+\eta_1^*}}{(k_2+k_1^*)} & \frac{e^{\eta_2+\xi_2^*}}{(k_2+l_2^*)} \\ 
	\frac{e^{\eta_1+\eta_1^*}}{(k_1+k_1^*)} & \frac{e^{\eta_1+\eta_2^*}}{(k_1+k_2^*)} & \frac{e^{\eta_1+\eta_1^*}}{(k_1+k_1^*)} & \frac{e^{\eta_1+\xi_2^*}}{(k_1+l_2^*)} \\ 
	\frac{e^{\xi_2+\eta_1^*}}{(l_2+k_1^*)} & \frac{e^{\xi_2+\eta_2^*}}{(l_2+k_2^*)} & \frac{e^{\xi_2+\eta_1^*}}{(l_2+k_1^*)} & \frac{e^{\xi_2+\xi_2^*}}{(l_2+l_2^*)} 
	\end{pmatrix},~\nonumber\\
	&&B=\begin{pmatrix}
	\frac{a\alpha_1^{(1)}\alpha_1^{(1)*}}{(k_1+k_1^*)} & \frac{a\alpha_2^{(1)}\alpha_1^{(1)*}}{(k_2+k_1^*)} & \frac{b^*\alpha_1^{(2)}\alpha_1^{(1)*}}{(k_1+k_1^*)}&\frac{b^*\alpha_2^{(2)}\alpha_1^{(1)*}}{(l_2+k_1^*)}\\ 
	\frac{a\alpha_1^{(1)}\alpha_2^{(1)*}}{(k_1+k_2^*)} & \frac{a\alpha_2^{(1)}\alpha_2^{(1)*}}{(k_2+k_2^*)} & \frac{b^*\alpha_1^{(2)}\alpha_2^{(1)*}}{(k_1+k_2^*)}&\frac{b^*\alpha_2^{(2)}\alpha_2^{(1)*}}{(l_2+k_2^*)} \\ 
	\frac{b\alpha_1^{(1)}\alpha_1^{(2)*}}{(k_1+k_1^*)} & \frac{b\alpha_2^{(1)}\alpha_1^{(2)*}}{(k_2+k_1^*)} & \frac{c\alpha_1^{(2)}\alpha_1^{(2)*}}{(k_1+k_1^*)}&\frac{c\alpha_2^{(2)}\alpha_1^{(2)*}}{(l_2+k_1^*)}\\
	\frac{b\alpha_1^{(1)}\alpha_2^{(2)*}}{(k_1+l_2^*)} & \frac{b\alpha_2^{(1)}\alpha_2^{(2)*}}{(k_2+l_2^*)} & \frac{c\alpha_1^{(2)}\alpha_2^{(2)*}}{(k_1+l_2^*)}&\frac{c\alpha_2^{(2)}\alpha_2^{(2)*}}{(l_2+l_2^*)}
	\end{pmatrix},\nonumber\\
	&&\phi=\begin{pmatrix}
	e^{\eta_1} & e^{\eta_2}  & e^{\eta_1}  & e^{\xi_2} 
	\end{pmatrix}^T. \label{3.10}
	\end{eqnarray} 

The structural and the interaction properties associated with this interesting class of solution are described by seven complex parameters $k_j$, $l_2$, $\alpha_1^{(j)}$, and $\alpha_2^{(j)}$, $j=1,2$.    

Furthermore, when we consider further restriction on the wave numbers, $k_1=l_1$ and $k_2=l_2$ we are able to capture the already known completely degenerate two-soliton solution \cite{senthil}. To bring out this solution through the Hirota method one has to assume the seed solutions as \begin{eqnarray}
	g_1^{(j)}=\al_1^{(j)}e^{\eta_1}+\al_2^{(j)}e^{\eta_2}, ~\eta_{j}=k_{j}t+ik_{j}^{2}z,~ j=1,2.~~~\label{11}
\end{eqnarray}
 Once again the final form of the pure degenerate two-soliton solution is same as the one presented in Eq. (\ref{3.7}) except now the matrices $A$, $B$, and $\phi$ take the following forms:
%%%%%%%%%%%%%%%%%%%%%%%%%%%%%%%%%

% \begin{widetext}
 	\begin{eqnarray}
 		&&A =
 		\begin{pmatrix}
 			\frac{e^{\eta_1+\eta_1^*}}{(k_1+k_1^*)} & \frac{e^{\eta_1+\eta_2^*}}{(k_1+k_2^*)} & \frac{e^{\eta_1+\eta_1^*}}{(k_1+k_1^*)} & \frac{e^{\eta_1+\eta_2^*}}{(k_1+k_2^*)} \\ 
 			\frac{e^{\eta_2+\eta_1^*}}{(k_2+k_1^*)} & \frac{e^{\eta_2+\eta_2^*}}{(k_2+k_2^*)} & \frac{e^{\eta_2+\eta_1^*}}{(k_2+k_1^*)} & \frac{e^{\eta_2+\eta_2^*}}{(k_2+k_2^*)} \\ 
 			\frac{e^{\eta_1+\eta_1^*}}{(k_1+k_1^*)} & \frac{e^{\eta_1+\eta_2^*}}{(k_1+k_2^*)} & \frac{e^{\eta_1+\eta_1^*}}{(k_1+k_1^*)} & \frac{e^{\eta_1+\eta_2^*}}{(k_1+k_2^*)} \\ 
 			\frac{e^{\eta_2+\eta_1^*}}{(k_2+k_1^*)} & \frac{e^{\eta_2+\eta_2^*}}{(k_2+k_2^*)} & \frac{e^{\eta_2+\eta_1^*}}{(k_2+k_1^*)} & \frac{e^{\eta_2+\eta_2^*}}{(k_2+l_2^*)} 
 		\end{pmatrix},~\nonumber\\
 		&&B=\begin{pmatrix}
 			\frac{a\alpha_1^{(1)}\alpha_1^{(1)*}}{(k_1+k_1^*)} & \frac{a\alpha_2^{(1)}\alpha_1^{(1)*}}{(k_2+k_1^*)} & \frac{b^*\alpha_1^{(2)}\alpha_1^{(1)*}}{(k_1+k_1^*)}&\frac{b^*\alpha_2^{(2)}\alpha_1^{(1)*}}{(k_2+k_1^*)}\\ 
 			\frac{a\alpha_1^{(1)}\alpha_2^{(1)*}}{(k_1+k_2^*)} & \frac{a\alpha_2^{(1)}\alpha_2^{(1)*}}{(k_2+k_2^*)} & \frac{b^*\alpha_1^{(2)}\alpha_2^{(1)*}}{(k_1+k_2^*)}&\frac{b^*\alpha_2^{(2)}\alpha_2^{(1)*}}{(k_2+k_2^*)} \\ 
 			\frac{b\alpha_1^{(1)}\alpha_1^{(2)*}}{(k_1+k_1^*)} & \frac{b\alpha_2^{(1)}\alpha_1^{(2)*}}{(k_2+k_1^*)} & \frac{c\alpha_1^{(2)}\alpha_1^{(2)*}}{(k_1+k_1^*)}&\frac{c\alpha_2^{(2)}\alpha_1^{(2)*}}{(k_2+k_1^*)}\\
 			\frac{b\alpha_1^{(1)}\alpha_2^{(2)*}}{(k_1+k_2^*)} & \frac{b\alpha_2^{(1)}\alpha_2^{(2)*}}{(k_2+k_2^*)} & \frac{c\alpha_1^{(2)}\alpha_2^{(2)*}}{(k_1+k_2^*)}&\frac{c\alpha_2^{(2)}\alpha_2^{(2)*}}{(k_2+k_2^*)}
 		\end{pmatrix},\nonumber\\
 		&&\phi=\begin{pmatrix}
 			e^{\eta_1} & e^{\eta_2}  & e^{\eta_1}  & e^{\eta_2} 
 		\end{pmatrix}^T. \label{12}
 \end{eqnarray} %\end{widetext} 
 
From the above one can observe that the structural and collisional behaviors associated with the degenerate two-solitons are governed by six complex parameters $k_j$, $\al_1^{(j)}$, and $\al_2^{(j)}$, $j=1,2$. Obviously, the degenerate two-soliton  (and $N$-soliton as well) solution of the GCNLS system (\ref{1}) is a sub-case of the nondegenerate two-soliton ($N$-soliton) solution (\ref{3.7}). We wish to remark that one can bring out the nondegenerate fundamental soliton solution (\ref{3.3a})-(\ref{3.3b}) from the degenerate two-soliton solution (solution (\ref{3.7}) with Eq. (12)) by setting $\al_2^{(1)}=0$ and $\al_1^{(2)}=0$ in the corresponding seed solution and renaming the complex constants $k_2$ and $\al_2^{(2)}$ as $l_1$ and $\al_1^{(2)}$, respectively. However, to the best of our knowledge such a choice and the resultant form of solution have not been discussed earlier in the literature. 
%%%%%%%%%%%%%%%%%%%%%%%%%%%%%%%%%%%%%%%%%%%%%%%%%%%%%
\section{Collision properties of nondegenerate solitons}
\begin{figure*}[t]
\centering
\includegraphics[width=0.9\linewidth]{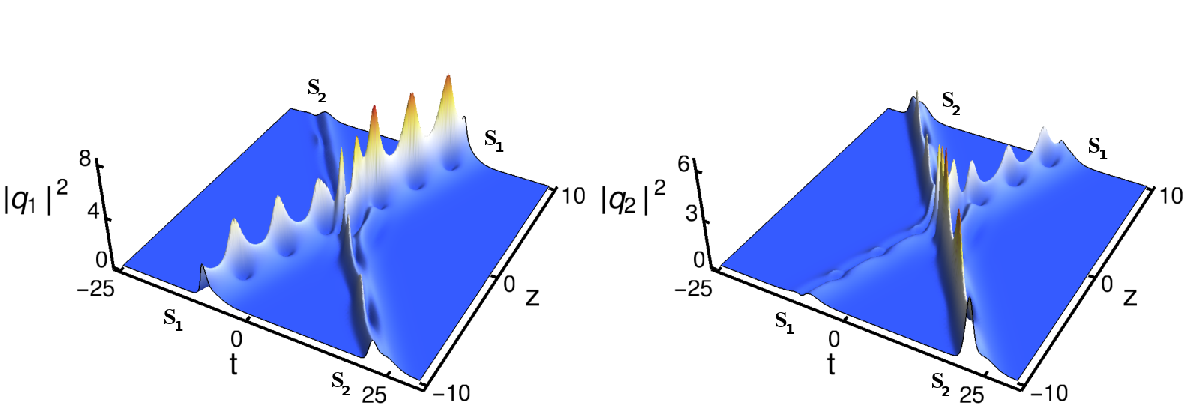}
\caption{Interaction between two breathing nondegenerate soliton states is demonstrated by fixing the parameter values as $a=c=1$, $b=0.5+0.5i$, $k_{1}=1.5+0.5i$, $l_{1}=0.5+0.5i$, $k_{2}=0.5-i$, $l_{2}=1.3-i$, $\alpha_{1}^{(1)}=0.5$, $\alpha_{2}^{(1)}=0.5+0.5i$, $\alpha_{1}^{(2)}=0.45+0.45i$ and $\alpha_{2}^{(2)}=1+i$. }
\label{f7}
\end{figure*}

Now, we intend to explore the collision behaviors of the nondegenerate solitons of the GCNLS system (\ref{1}) with respect to the following two cases: (i) strong FWM effect ($b\ge0.5+0.5i$), and (ii) weak FWM effect ($b\ge0.15+0.15i$) along with further choices of the real parts of the wave numbers $k_j$ and $l_j$, $j=1,2$. Here, we restrict ourselves to the equal velocities case, $k_{1I}=l_{1I}$, and $k_{2I}=l_{2I}$. For both strong and weak FWM effects, the two nondegenerate solitons undergo novel shape changing collisions and shape preserving collisions for suitable choices of real parts of the wave numbers. These collision scenarios associated with the nondegenerate solitons can be analyzed by deducing the asymptotic forms of them from the two-soliton solution (\ref{3.7}) of Eq. (\ref{1}).   Here, the asymptotic analysis corresponding to the shape changing collisions with strong FWM effect (Figs. \ref{f7} and \ref{f8}) is only presented and it can be carried out for shape preserving case and for other cases also in a similar manner.   
\subsection{Asymptotic analysis}
We study the interaction dynamics of the nondegenerate solitons by deducing the explicit forms of the individual solitons from the nondegenerate two-soliton solution (\ref{3.7}) at the limits $z\rightarrow \pm\infty$. To deduce these, we consider $k_{1I}>k_{2I}$, $l_{1I}>l_{2I}$, $k_{jR},l_{jR}>0$, $j=1,2$,   $k_{1I}=l_{1I}$ and $k_{2I}=l_{2I}$ as typical examples. The latter parametric choices correspond to head-on collision among the two nondegenerate solitons. In this circumstance, the two nondegenerate solitons $S_1$ and $S_2$ are well separated and subsequently their asymptotic forms can be deduced from the solution (\ref{3.7}) by considering the asymptotic nature of the wave variables $\xi_{jR}=l_{jR}(t-2l_{jI}z)$, and  $\eta_{jR}=k_{jR}(t-2k_{jI}z)$, $j=1,2$. The asymptotic behavior of the variables $\eta_{jR}$ and $\xi_{jR}$ are obtained as (i) Soliton 1 ($S_1$): $\eta_{1R}$, $\xi_{1R}\simeq 0$, $\eta_{2R}$, $\xi_{2R}\rightarrow\mp \infty$ as $z\rightarrow\mp\infty$ and (ii) Soliton 2 ($S_2$): $\eta_{2R}$, $\xi_{2R}\simeq 0$, $\eta_{1R}$, $\xi_{1R}\rightarrow\mp \infty$ as $z\rightarrow\pm\infty$. These results lead to the following asymptotic forms of the individual nondegenerate solitons. Note that the various phase constants appearing in the following asymptotic expressions are defined in \ref{B} for convenience.\\
%\begin{widetext}\vspace{0.21cm}
(a) Before collision: $z\rightarrow -\infty$\\
{\bf Soliton 1}: $\eta_{1R},~\xi_{1R}\simeq0$,  $\eta_{2R},~\xi_{2R}\rightarrow-\infty$\\
In this asymptotic limit, the forms of $q_1$ and $q_2$ are deduced from the two-soliton solution (\ref{3.7}) for soliton 1 as below:
\bes\bea
&&\hspace{-1.0cm}q_1=\frac{1}{D_1^-}\big(e^{i\eta_{1I}}c_{11}^{1-}\cosh(\xi_{1R}+\phi_1^{1-})+c_{12}^{1-}e^{i\xi_{1I}}[\cosh\eta_{1R}+\sinh\eta_{1R}]\big),\label{13a}\\
&&\hspace{-1.0cm}q_2=\frac{1}{D_1^-}\big(e^{i\xi_{1I}}c_{21}^{1-}\cosh(\eta_{1R}+\phi_2^{1-})+c_{22}^{1-}e^{i\eta_{1I}}[\cosh\xi_{1R}+\sinh\xi_{1R}]\big),\label{13b}\\
&&\hspace{-1.0cm}D_1^-=\Lam_1^{1-}\cosh(\eta_{1R}+\xi_{1R}+\phi_3^{1-})+\Lam_2^{1-}\cosh(\eta_{1R}-\xi_{1R}+\phi_4^{1-})\nonumber\\
&&\hspace{0.0cm}+\Lam_3^{1-}\big[\cosh\phi_5^{1-}\cos(\eta_{1I}-\xi_{1I})+i\sinh\phi_5^{1-}\sin(\eta_{1I}-\xi_{1I})\big],\nonumber
\eea\ees
where $c_{11}^{1-}=e^{\frac{\ga_2+\rho_1}{2}}$, $c_{12}^{1-}=\frac{1}{2}e^{\ga_1}$,  $c_{21}^{1-}=e^{\frac{\nu_1+\rho_2}{2}}$, $c_{22}^{1-}=\frac{1}{2}e^{\nu_2}$,  $\phi_1^{1-}=\frac{\ga_2-\rho_1}{2}$, $\phi_2^{1-}=\frac{\nu_1-\rho_2}{2}$, $\phi_3^{1-}=\frac{\lam_1}{2}$, $\phi_4^{1-}=\frac{\del_1-\del_{13}}{2}$, $\phi_5^{1-}=\frac{\del_5-\del_6}{2}$, $\Lam_1^{1-}=e^{\frac{\lam_1}{2}}$, $\Lam_2^{1-}=e^{\frac{\del_1+\del_{13}}{2}}$, $\Lam_3^{1-}=e^{\frac{\del_5+\del_6}{2}}$, and $\rho_j=\log\al_1^{(j)}$, $j=1,2$. Here, the superscript $1-$ denotes the soliton 1 before collision.\\ 
%\bes
%\bea
%&&q_1=\frac{1}{D_1^-}\big(\alpha_1^{(1)}e^{\eta_1}+e^{\eta_1+\eta_1^*+\xi_1+\ga_1}+e^{\eta_1+\xi_1+\xi_1^*+\ga_2}\big),\label{13a}\\
%&&q_2=\frac{1}{D_1^-}\big(\alpha_1^{(2)}e^{\xi_1}+e^{\eta_1+\eta_1^*+\xi_1+\nu_1}+e^{\eta_1+\xi_1+\xi_1^*+\nu_2}\big),\label{13b}\\
%&&D_1^-=1+e^{\eta_1+\eta_1^*+\del_1}+e^{\eta_1+\xi_1^*+\del_5}+e^{\eta_1^*+\xi_1+\del_6}+e^{\xi_1+\xi_1^*+\del_{13}}+e^{\eta_1+\eta_1^*+\xi_1+\xi_1^*+\lam_1}.\nonumber
%\eea
%\ees
{\bf Soliton 2}: $\eta_{2R},~\xi_{2R}\simeq0$,  
$\eta_{1R},~\xi_{1R}\rightarrow+\infty$\\
In this limit, the asymptotic forms for soliton 2 are deduced as follows:
\bes
\begin{eqnarray}
	&&q_1=\frac{1}{D_2^-}\big(e^{i\eta_{2I}}c_{11}^{2-}\cosh(\xi_{2R}+\phi_1^{2-})+e^{i\xi_{2I}}c_{12}^{2-}\cosh(\eta_{2R}+\phi_6^{2-})\big),\label{14a}\\
	&&q_2=\frac{1}{D_2^-}\big(e^{i\eta_{2I}}c_{22}^{2-}\cosh(\xi_{2R}+\phi_7^{2-})+e^{i\xi_{2I}}c_{21}^{2-}\cosh(\eta_{2R}+\phi_2^{2-})\big),\label{14b}\\
	&&D_2^-=\Lam_1^{2-}\cosh(\eta_{2R}+\xi_{2R}+\phi_3^{2-})+\Lam_2^{2-}\cosh(\eta_{2R}-\xi_{2R}+\phi_4^{2-})\nonumber\\
	&&\hspace{1.0cm}+\Lam_3^{2-}\big[\cosh\phi_5^{2-}\cos(\eta_{2I}-\xi_{2I})+i\sinh\phi_5^{2-}\sin(\eta_{2I}-\xi_{2I})\big],\nonumber
\end{eqnarray}
\ees
where $c_{11}^{2-}=e^{\frac{\mu_{28}+\mu_1}{2}}$, $c_{12}^{2-}=e^{\frac{\mu_{27}+\mu_4}{2}}$, $c_{21}^{2-}=e^{\frac{\chi_{26}+\chi_4}{2}}$, $c_{22}^{2-}=e^{\frac{\chi_{27}+\chi_1}{2}}$, $\Lam_1^{2-}=e^{\frac{r_{17}+\lam_1}{2}}$, $\Lam_2^{2-}=e^{\frac{r_{1}+r_8}{2}}$, $\Lam_3^{2-}=e^{\frac{r_{2}+r_5}{2}}$, $\phi_1^{2-}=\frac{\mu_{28}-\mu_1}{2}$,  $\phi_2^{2-}=\frac{\chi_{26}-\chi_4}{2}$, $\phi_3^{2-}=\frac{r_{17}-\lam_1}{2}$, $\phi_4^{2-}=\frac{r_{1}-r_8}{2}$,  $\phi_5^{2-}=\frac{r_{2}-r_5}{2}$, $\phi_6^{2-}=\frac{\mu_{27}-\mu_4}{2}$, and $\phi_7^{2-}=\frac{\chi_{27}-\chi_1}{2}$. In the latter, the superscript $2-$ denotes the soliton 2 before collision.\\ 
%\bes
%\bea
%&&q_1=\frac{1}{D_2^-}\big(e^{\eta_2+\mu_1}+e^{\xi_2+\mu_4}+e^{\eta_2+\eta_2^*+\xi_2+\mu_{27}}+e^{\eta_2+\xi_2+\xi_2^*+\mu_{28}}\big),\label{14a}\\
%&&q_2=\frac{1}{D_2^-}\big(e^{\eta_2+\chi_1}+e^{\xi_2+\chi_4}+e^{\eta_2+\eta_2^*+\xi_2+\chi_{26}}+e^{\eta_2+\xi_2+\xi_2^*+\chi_{27}}\big),\label{14b}
%\\
%&&D_2^-=e^{\lam_1}+e^{\eta_2+\eta_2^*+r_1}+e^{\eta_2+\xi_2^*+r_5}+e^{\eta_2^*+\xi_2+r_2}+e^{\xi_2+\xi_2^*+r_8}+e^{\eta_2+\eta_2^*+\xi_2+\xi_2^*+r_{17}}.\nonumber
%\eea
%\ees	
(b) After collision: $z\rightarrow +\infty$\\
{\bf Soliton 1}: $\eta_{1R},~\xi_{1R}\simeq0$,  
$\eta_{2R},~\xi_{2R}\rightarrow+\infty$\\
As we mentioned above, the asymptotic forms corresponding to the soliton 1 after collision can also be deduced from the two-soliton solution (\ref{3.7}) and they read as
\bes
\begin{eqnarray}
	&&q_1=\frac{1}{D_1^+}\big(e^{i\eta_{1I}}c_{11}^{1+}\cosh(\xi_{1R}+\phi_1^{1+})+e^{i\xi_{1I}}c_{12}^{1+}\cosh(\eta_{1R}+\phi_6^{1+})\big),\label{15a}\\
	&&q_2=\frac{1}{D_1^+}\big(e^{i\eta_{1I}}c_{22}^{1+}\cosh(\xi_{1R}+\phi_7^{1+})+e^{i\xi_{1I}}c_{21}^{1+}\cosh(\eta_{1R}+\phi_2^{1+})\big),\label{15b}\\
	&&D_1^+=\Lam_1^{1+}\cosh(\eta_{1R}+\xi_{1R}+\phi_5^{1+})+\Lam_2^{1+}e\cosh(\eta_{1R}-\xi_{1R}+\phi_6^{1+})\nonumber\\
	&&\hspace{1.0cm}+\Lam_3^{1+}\big[\cosh\phi_7^{1+}\cos(\eta_{1I}-\xi_{1I})+i\sinh\phi_7^{1+}\sin(\eta_{1I}-\xi_{1I})\big],\nonumber
\end{eqnarray}
\ees
where $c_{11}^{1+}=e^{\frac{\mu_{26}+\mu_{23}}{2}}$, $c_{12}^{1+}=e^{\frac{\mu_{25}+\mu_{24}}{2}}$, $c_{21}^{1+}=e^{\frac{\chi_{28}+\chi_{24}}{2}}$, $c_{22}^{1+}=e^{\frac{\chi_{25}+\chi_{23}}{2}}$, $\Lam_1^{1+}=e^{\frac{r_{17}+\lam_{36}}{2}}$, $\Lam_2^{1+}=^{\frac{r_{13}+r_{16}}{2}}$, $\Lam_3^{1+}=e^{\frac{r_{14}+r_{15}}{2}}$, $\phi_1^{1+}=\frac{\mu_{26}-\mu_{23}}{2}$,  $\phi_2^{1+}=\frac{\chi_{28}-\chi_{24}}{2}$, $\phi_3^{1+}=\frac{r_{17}-\lam_{36}}{2}$, $\phi_4^{1+}=\frac{r_{13}-r_{16}}{2}$, $\phi_5^{1+}=\frac{r_{14}-r_{15}}{2}$, $\phi_6^{1+}=\frac{\mu_{25}-\mu_{24}}{2}$, and $\phi_7^{1+}=\frac{\chi_{25}-\chi_{23}}{2}$.  Here, the superscript $1+$ represents the soliton 1 after collision.\\
%\bes
%\bea
%&&q_1=\frac{1}{D_1^+}\big(e^{\eta_1+\mu_{23}}+e^{\xi_1+\mu_{24}}+e^{\eta_1+\eta_1^*+\xi_1+\mu_{25}}+e^{\eta_1+\xi_1+\xi_1^*+\mu_{26}}\big),\label{15a}\eea\bea
%&&q_2=\frac{1}{D_1^+}\big(e^{\eta_1+\chi_{23}}+e^{\xi_1+\chi_{24}}+e^{\eta_1+\eta_1^*+\xi_1+\chi_{25}}+e^{\eta_1+\xi_1+\xi_1^*+\chi_{28}}\big),\label{15b}\\
%&&D_1^+=e^{\lam_{36}}+e^{\eta_1+\eta_1^*+r_{13}}+e^{\eta_1+\xi_1^*+r_{15}}+e^{\eta_1^*+\xi_1+r_{14}}+e^{\xi_1+\xi_1^*+r_{16}}+e^{\eta_1+\eta_1^*+\xi_1+\xi_1^*+r_{17}}.\nonumber
%\eea
%\ees	
{\bf Soliton 2}: $\eta_{2R},~\xi_{2R}\simeq0$,  $\eta_{1R}, ~\xi_{1R}\rightarrow-\infty$\\
Similarly, we have obtained the asymptotic forms of $q_1$ and $q_2$ from the two soliton solution (\ref{3.7}) for soliton 2 as below:
\bes\bea
&&\hspace{-1.0cm}q_1=\frac{1}{D_2^+}\big(e^{i\eta_{2I}}c_{11}^{2+}\cosh(\xi_{2R}+\phi_1^{2+})+c_{12}^{2+}e^{i\xi_{2I}}[\cosh\eta_{2R}+\sinh\eta_{2R}]\big),\label{16a}\\
&&\hspace{-1.0cm}q_2=\frac{1}{D_2^+}\big(e^{i\xi_{2I}}c_{21}^{2+}\cosh(\eta_{2R}+\phi_2^{2+})+e^{i\eta_{2I}}c_{22}^{2+}[\cosh\xi_{2R}+\sinh\xi_{2R}]\big),\label{16b}\\
&&\hspace{-1.0cm}D_2^+=\Lam_1^{2+}\cosh(\eta_{2R}+\xi_{2R}+\phi_3^{2+})+\Lam_2^{2+}\cosh(\eta_{2R}-\xi_{2R}+\phi_4^{2+})\nonumber\\
&&\hspace{0.0cm}+\Lam_3^{2+}\big[\cosh\phi_5^{2+}\cos(\eta_{2I}-\xi_{2I})+i\sinh\phi_5^{2+}\sin(\eta_{2I}-\xi_{2I})\big],\nonumber
\eea\ees
where $c_{11}^{2+}=e^{\frac{\ga_{20}+\rho_1'}{2}}$, $c_{12}^{2+}=\frac{1}{2}e^{\ga_{15}}$, $c_{21}^{2+}=e^{\frac{\nu_{15}+\rho_2'}{2}}$, $c_{22}^{2+}=\frac{1}{2}e^{\nu_{20}}$, $\Lam_1^{2+}=e^{\frac{\lam_{36}}{2}}$, $\Lam_2^{2+}=e^{\frac{\del_4+\del_{16}}{2}}$, $\Lam_3^{2+}=e^{\frac{\del_{11}+\del_{12}}{2}}$, $\phi_1^{2+}=\frac{\ga_{20}-\rho_1'}{2}$, $\phi_2^{2+}=\frac{\nu_{15}-\rho_2'}{2}$, $\phi_3^{2+}=\frac{\lam_{36}}{2}$, $\phi_4^{2+}=\frac{\del_4-\del_{16}}{2}$, $\phi_5^{2+}=\frac{\del_{11}-\del_{12}}{2}$, $\rho_j'=\log\al_2^{(j)}$, $j=1,2$.  Here, the superscript $2+$ represents the soliton 2 after collision. 
%\end{widetext}
From the above asymptotic expressions, one can distinguish the shape changing collisions from the shape preserving collision by calculating the constants, $c_{nm}^{l\pm}$, $\Lam_j^{l\pm}$, $n,m,l=1,2$,  $j=1,2,3$, and the phase terms, $\phi_{k}^{l\pm}$,  $k=1,2,3,4,5$, $\phi_{6,7}^{1+}$, and $\phi_{6,7}^{2-}$, explicitly. In general, these complex quantities are not preserved during the collision as it is true from their corresponding asymptotic forms. Because of this variation, the nondegenerate solitons, in general, undergo shape changing collision. However, the shape preserving collision always takes place whenever $c_{nm}^{l+}=c_{nm}^{l-}$, $\Lam_{j}^{l+}=\Lam_{j}^{l-}$, and $\phi_{k}^{l+}=\phi_{k}^{l-}$, otherwise the shape changing collision will occur. The occurrence of these collision scenarios mainly depends the real parts of the wave numbers $k_j$, and $l_j$, $j=1,2$. We note that these constants and the phase terms of solitons 1 and 2 before and after collision are related. However, here we have omitted these details because of the complex forms of the asymptotic expressions.
%%%%%%%%%%%%%%%%%%%%%%%%%%%%%%%%%%%%%%
%%%%%%%%%%%%%%%%%%%%%%%%%%%%%%%%%%%%%%%%%%%5%%%%
\subsection{Strong FWM effect: Shape changing and shape preserving collisions}
The above asymptotic analysis reveals that there is a definite possibility of observing shape changing collision among the two nondegenerate solitons since the asymptotic expressions (\ref{13a})-(\ref{13b}) of soliton 1 and (\ref{14a})-(\ref{14b}) of soliton 2 are not preserved after the collision process. In the present nondegenerate case, the shape changing that occurs is essentially due to the drastic variations in the phase terms, as it has been explained in the case of Manakov system \cite{ramakrishnan,stalin-review}, and because of the changes in the constants, $c_{nm}^{l\pm}$, $\Lam_j^{l\pm}$, $n,m,l=1,2$,  $j=1,2,3$, along with the FWM effect. We note here that the asymptotic expressions, with $b=0$ and $a=c=1$, given above coincide with the one that were already reported for the Manakov system \cite{ramakrishnan}, where the structures of the nondegenerate solitons are mainly influenced by the phases only. Then, another important feature that we observe from the present analysis is the appearance of periodic functions, $\cos(\eta_{jI}-\xi_{jI})$, $\sin(\eta_{jI}-\xi_{jI})$, $\eta_{jI}=k_{jI}t+(k_{jR}^2-k_{jI}^2)z$, $\xi_{jI}=l_{jI}t+(l_{jR}^2-l_{jI}^2)z$, $j=1,2$, in the denominators of the asymptotic expressions (\ref{13a})-(\ref{13b}), (\ref{14a})-(\ref{14b}), (\ref{15a})-(\ref{15b}), and (\ref{16a})-(\ref{16b}). It implies that, in general, the breathing nature will appear on the structures of the nondegenerate solitons before and after collision with enough strength of FWM effect. And also to bring out this breathing behavior one has to consider any one of the following choice of real parts of wave numbers: (i) $k_{jR}^2>l_{jR}^2$, (ii) $k_{jR}^2<l_{jR}^2$, $j=1,2$, (iii) $k_{1R}^2>l_{1R}^2$, $k_{2R}^2<l_{2R}^2$, and (iv) $k_{1R}^2<l_{1R}^2$, $k_{2R}^2>l_{2R}^2$. Under these conditions, there is a possibility of occurrence of the intensity enhancement or suppression in the breathing soliton states after the collision process, as it is evident from the asymptotic forms, along with a finite phase shift. A typical shape changing collision among the two oppositely propagating breathing nondegenerate soliton states is displayed in Fig. \ref{f7}. From this figure, one can observe that initially the two breathing solitons are well separated and they undergo head-on collision. As a consequence of this collision, the intensity of oscillations of the soliton 1 ($S_1$) gets enhanced in both the modes $q_1$ and $q_2$. On the other hand, in order to obey the energy conservation,
\begin{equation}
\int_{-\infty}^{+\infty}|q_j|^2dt=\text{constant},~j=1,2, \label{17}
\end{equation}
in the individual components, the intensity of the oscillation gets suppressed in the other soliton, say $S_2$, in both the modes. That is, for a given soliton (say $S_1$), the enhancement of energy occurs in both the modes. This interesting collision scenario essentially appears because of the presence of the phase dependent nonlinearity $(bq_{1}q_{2}^{*}+b^{*}q_{1}^{*}q_{2})q_j$, $j=1,2$, as well as the changes that occurred in the phase terms and in the constants, $c_{nm}^{l\pm}$, $\Lam_j^{l\pm}$, $n,m,l=1,2$,  $j=1,2,3$, of the asymptotic forms of the individual solitons. In this case, these constants vary their forms during the collision process. One can characterize this shape changing collision scenario by finding the variations in these constants and in the phases. In this situation, transition intensities ($|T^l_j|^2=\frac{|A_j^{l+}|^2}{|A_j^{l-}|^2}$, $l,j=1,2$), will not be unimodular.  Apart from this, the total energy of the solitons in both the modes is also conserved,
\begin{equation}
\frac{d}{dz}\int_{-\infty}^{+\infty}(|q_1|^2+|q_2|^2)dt=0.\label{18}
\end{equation} 
This kind of energy sharing collision is similar to the collision scenario of the degenerate bright solitons in the present GCNLS system (\ref{1}) \cite{senthil} as well as in the mixed CNLS system ($b=0$, $a=-c=1$ in Eq. (\ref{1})) \cite{mix}, where the given degenerate soliton experiences the same kind of effect (energy enhancement/suppression) in each component through intensity redistribution. 

An interesting fact that can be observed both from Fig. \ref{f7} and the asymptotic expressions of the two solitons, before and after collision, is the maintaining of uniform periodicity throughout the collision scenario. It means that the time period of oscillations, 
\begin{equation}
T^{\pm}_j=\frac{2\pi}{k_{jR}^2-l_{jR}^2},~j=1,2,
\end{equation} 
remains constant during the collision though the intensities of oscillations get changed. We remark here that this novel shape changing collision of nondegenerate vector solitons has not been observed earlier in the Manakov system \cite{radha,ramakrishnan} and is new to the literature. This type of soliton collision may be useful in soliton based signal amplification application where the nondegenerate soliton $S_2$ acts like a pump wave and the soliton $S_1$ acts as a signal wave.  
       
\begin{figure*}[t]
\centering
\includegraphics[width=0.9\linewidth]{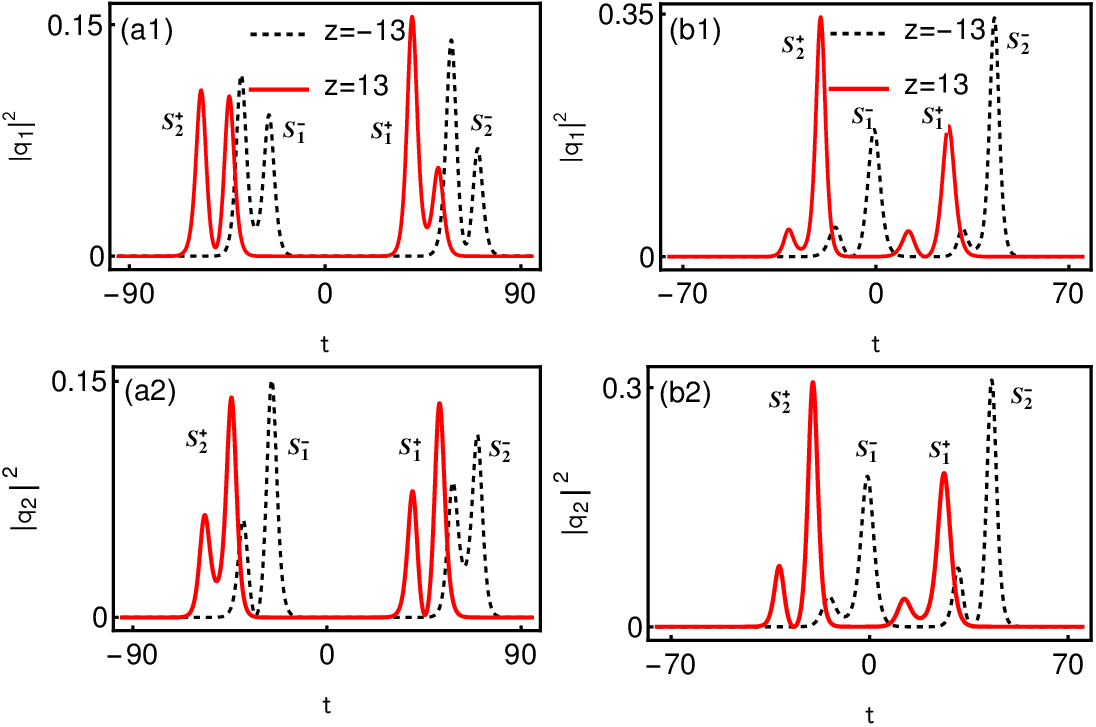}
\caption{The left panel demonstrates the shape changing collision among the two asymmetric double-hump solitons and the right panel illustrates the shape preserving collision among the two asymmetric double-hump nondegenerate solitons. To obtain Figs. (a1)-(a2) we fix the parameter values as $k_{1}=0.333+0.5i$, $l_{1}=0.315+0.5i$, $k_{2}=0.315-2.2i$, $l_{2}=0.333-2.2i$, $\alpha_{1}^{(1)}=\alpha_{2}^{(1)}=0.6$, $\alpha_{1}^{(2)}=\alpha_{2}^{(2)}=-0.45i$ whereas to draw Figs. (b1)-(b2), we consider the parameter values as $k_{1}=0.325+0.5i$, $l_{1}=0.35+0.5i$, $k_{2}=0.45-1.2i$, $l_{2}=0.425-1.2i$, $\alpha_{1}^{(1)}=0.5+0.5i$, $\alpha_{2}^{(1)}=0.5$, $\alpha_{1}^{(2)}=0.45+0.5i$, and $\alpha_{2}^{(2)}=0.5+0.5i$. }
\label{f8}
\end{figure*}

Further, it is also possible to observe the shape changing collision among the two non-breathing nondegenerate solitons, where the shape changing occurs in between the two asymmetric double-hump solitons. In this case, even for the strong FWM effect ($b=0.5+0.5i$), the shape changing property mainly relies on the appropriate choice of the real parts of the wave numbers, $k_j$, and  $l_j$, $j=1,2$. A typical shape changing collision among the two non-breathing asymmetric double-hump solitons is demonstrated in Fig. \ref{f8} (a1)-(a2) as an example.  From this figure, one can identify that the structures of initial set of asymmetric double-hump solitons before collision get changed into another set of asymmetric double-hump solitons. This structural deformation of the nondegenerate double-hump solitons essentially occurs because of the phase term variation as it is evident from the asymptotic phase forms. The phase terms of soliton 1 (soliton 2), $\phi_j^{1-}$ ($\phi_k^{2-}$), $j=1,2,3,4,5$, $k=1,2,...,7$, before collision get changed to $\phi_k^{1+}$ ($\phi_j^{2+}$) after collision. In this case also, the constants, $c_{nm}^{l\pm}$, $\Lam_j^{l\pm}$, $n,m,l=1,2$,  $j=1,2,3$, do not preserve their forms and they contribute to the shape changing nature of the nondegenerate solitons.        
%%%%%%%%%%%%%%%%%%%%%%%%%%%%%%%%%%%%%%%%%%%%%%%
%%%%%%%%%%%%%%%%%%%%%%%%%%%%%%%%%%%%%%%%%%%%%%%

Furthermore, in the present GCNLS system (\ref{1}), the nondegenerate solitons can also exhibit the shape preserving collision for a special choice of wave numbers. To observe this collision scenario, the constants, $c_{nm}^{l\pm}$, $\Lam_j^{l\pm}$, $n,m,l=1,2$,  $j=1,2,3$, should preserve their forms and the phase terms do not contribute to changing the structures of the nondegenerate solitons, as it has been pointed out in the Manakov case \cite{stalin-review}, thereby leading to an elastic collision. Such shape preserving collision is depicted in Fig. \ref{f8}(a2)-(b2), in which the two nondegenerate solitons can pass through each other without experiencing a phase shift. One can derive the zero phase shift criterion \cite{stalin-review,stalin-lsri} from the asymptotic expressions of individual solitons by finding the relations between the phase terms at $z\rightarrow\pm\infty$. For brevity, we have omitted the details due to the complex nature of analytical expressions.  
%%%%%%%%%%%%%%%%%%%%%%%%%%%%%%%%%%%%%%
\begin{figure*}[t]
\centering
\includegraphics[width=0.95\linewidth]{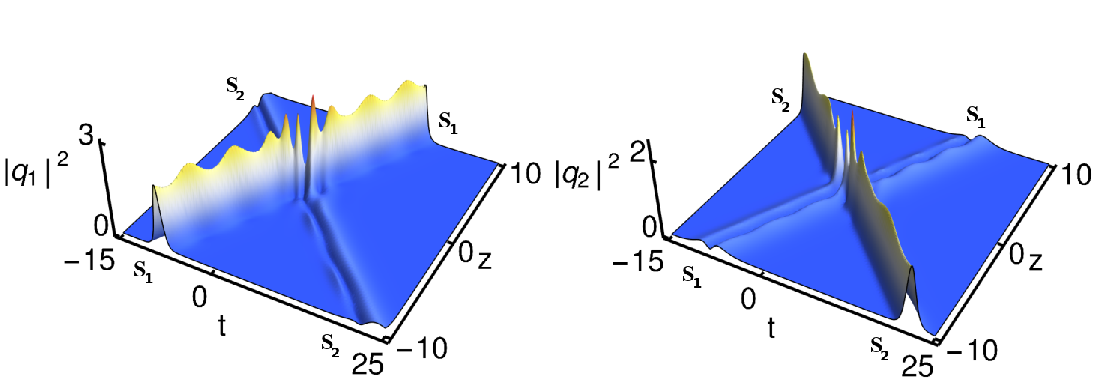}
\caption{An elastic collision among the two weakly breathing nondegenerate soliton states is demonstrated for $a=c=1$, $k_{1}=1.5+0.5i$, $l_{1}=0.45+0.5i$, $k_{2}=0.5-i$, $l_{2}=1.3-i$, $\alpha_{1}^{(1)}=0.5$, $\alpha_{2}^{(1)}=0.5+0.5i$, $\alpha_{1}^{(2)}=0.45+0.45i$ and $\alpha_{2}^{(2)}=1+i$. Here, the two solitons interact almost elastically and their structures remain preserved throughout the collision process.  }
\label{f9}
\end{figure*}
%%%%%%%%%%%%%%%%%%%%%%%%%%%%%%%%%%%%%%%%%%%%%%%
\begin{figure*}[t]
\centering
\includegraphics[width=0.9\linewidth]{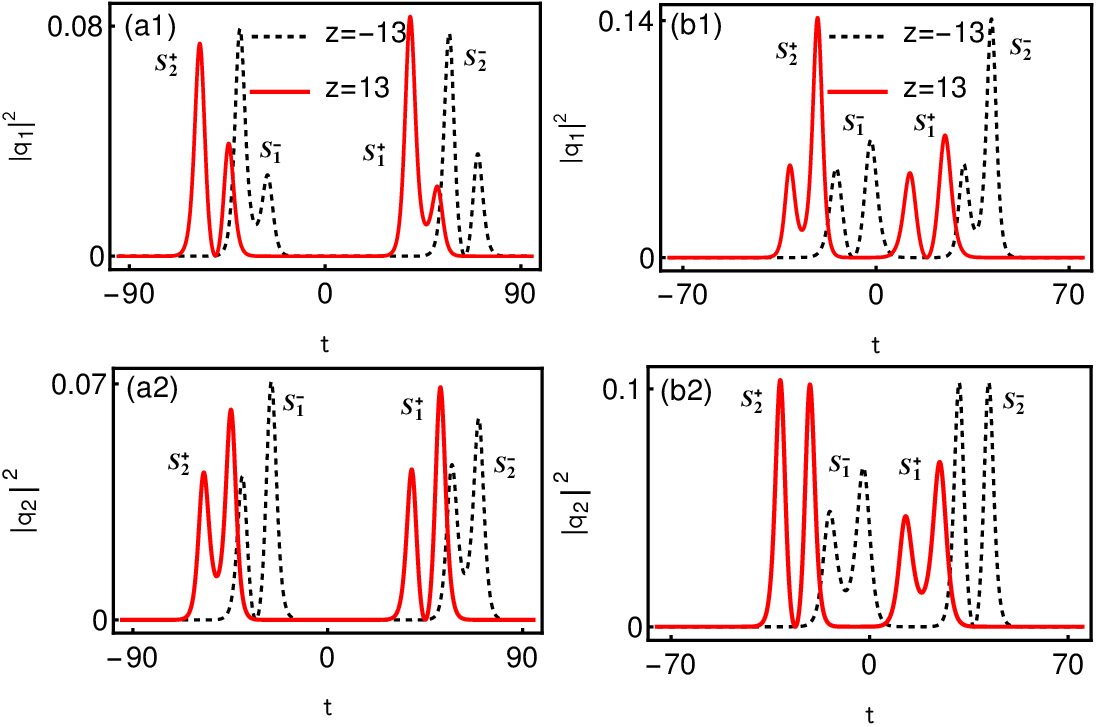}
\caption{The left panel represents the shape changing collision scenario among the two asymmetric double-hump solitons by fixing the parameter values as $a=c=1$, $k_{1}=0.333-0.5i$, $l_{1}=0.315-0.5i$, $k_{2}=0.315+2.2i$, $l_{2}=0.333+2.2i$, $\alpha_{1}^{(1)}=0.45+0.45i$, $\alpha_{2}^{(1)}=2.49+2.45i$, $\alpha_{1}^{(2)}=0.49+0.45i$ and $\alpha_{2}^{(2)}=0.45+0.45i$. The right panel illustrates the shape preserving collision scenario among the two asymmetric double-hump solitons. To display this collision scenario we fix the parameter values as $a=c=1$, $k_{1}=0.325+0.5i$, $l_{1}=0.35+0.5i$, $k_{2}=0.45-1.2i$, $l_{2}=0.425-1.2i$, $\alpha_{1}^{(1)}=0.5+0.5i$, $\alpha_{2}^{(1)}=0.5$, $\alpha_{1}^{(2)}=0.45+0.5i$ and $\alpha_{2}^{(2)}=0.5+0.5i$. }
\label{f10}
\end{figure*}
%%%%%%%%%%%%%%%%%%%%%%%%%%%%%%%%%%%%%%%%%%%
\subsection{Weak FWM effect: Shape preserving and shape changing collisions}
To understand the collision scenario of nondegenerate vector solitons in the presence of weak FWM effect, we consider the choice of wave numbers as $k_{1I}>k_{2I}$, $l_{1I}>l_{2I}$, $k_{jR},l_{jR}>0$, $j=1,2$. The latter condition on the wave numbers is the same as the one fixed earlier to analyze the effect of strong FWM effect on the interaction among the nondegenerate solitons. Therefore, using the same asymptotic analysis that presented in Section III A one will be able to understand the effect of weak FWM on the collision dynamics of nondegenerate solitons. To analyze this, now we fix the FWM parameter value $b$ as $0.15+0.15i$. In this circumstance, the nondegenerate solitons with weak breathing property exhibit a mere shape preserving collision as it is illustrated in Fig. \ref{f9} for $a=c=1$, $k_{1}=1.5+0.5i$, $l_{1}=0.45+0.5i$, $k_{2}=0.5-i$, $l_{2}=1.3-i$, $\alpha_{1}^{(1)}=0.5$, $\alpha_{2}^{(1)}=0.5+0.5i$, $\alpha_{1}^{(2)}=0.45+0.45i$ and $\alpha_{2}^{(2)}=1+i$. From this figure, one can infer that the two weakly breathing nondegenerate solitons interact almost elastically with a slight phase shift. It means that the structures of the two solitons remain constant and subsequently they pass through each other with almost a zero phase shift. In this situation, the phase dependent nonlinearity, $(bq_{1}q_{2}^{*}+b^{*}q_{1}^{*}q_{2})q_j$, $j=1,2$, plays less role in affecting the collision dynamics of solitons. This kind of shape preserving nature of non-degenerate solitons will be useful in optical telecommunication applications, where the input signal does not distort during propagation in fibers.          

On the other hand, very interestingly we also observe that the two asymmetric double-hump nondegenerate solitons with no breathing behavior undergo a non-trivial shape changing collision (but without energy exchange), as it is demonstrated in Fig. \ref{f10}(a1)-(a2), even for the low strength of FWM. From this figure, one can find that the asymmetric double-hump solitons lose their original structures and they become another set of asymmetric double-hump solitons as a final product of the collision scenario. This type of collision essentially arises due to the changes in the phase terms. Apart from the above, the nondegenerate solitons exhibit almost shape preserving collision, as it is demonstrated in Fig. \ref{f10} (b1)-(b2), again for low strengths of FWM effect. In this situation, the shapes of the two asymmetric double-hump solitons remain almost invariant under collision thereby confirming the elastic collision nature. The elastic nature of the collision scenario can be confirmed by calculating the transition intensities from the asymptotic forms, where the phase terms do not vary throughout the collision scenario. 
   
%\begin{eqnarray}
%\kappa_{mm'}=\frac{\psi_m^{\dagger}\sigma\psi_{m'}}{(k_m^*+k_{m'})},~\kappa_{mn}=\frac{\psi_m^{\dagger}\sigma\psi'_{n}}{(k_m^*+l_{n})},\nonumber\\
%\kappa_{nm}=\frac{\psi_n^{'\dagger}\sigma\psi_{m}}{(l_n^*+k_{m})},~\kappa_{nn'}=\frac{\psi_n^{'\dagger}\sigma\psi'_{n'}}{(l_n^*+l_{n'})}.\nonumber
%\end{eqnarray} 

%\begin{equation*}		
%\kappa_{nm} = %\frac{\Psi^{'\dagger}_n\sigma\Psi_m}{(l^*_n+k_m)}, ~ %\kappa_{nn} = %\frac{\Psi^{'\dagger}_m\sigma\Psi_{n'}^'}{(l^*_n+l_{n'})}.	
%\end{equation*}

%%%%%%%%%%%%%%%%%%%%%%%%%%%%%%%%%%%%%%%%%%%%%%%
\begin{figure*}[t]
\centering
\includegraphics[width=0.95\linewidth]{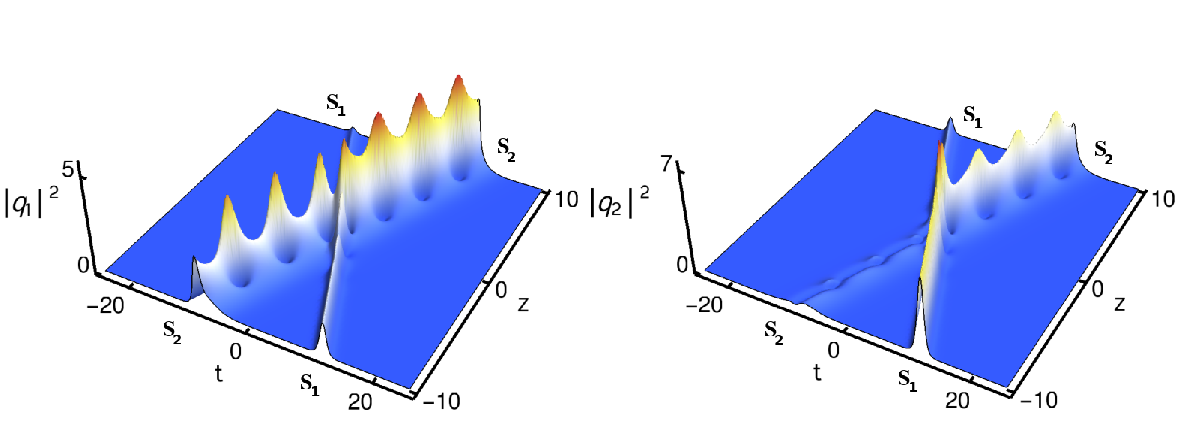}
\caption{Type-I energy sharing collision between the degenerate and nondegenerate solitons is illustrated by fixing the parameter values as $a=c=1$, $k_{1}=l_1=1.5-0.5i$, $k_{2}=1.5+0.5i$, $l_{2}=0.45+0.5i$, $\alpha_{1}^{(1)}=0.5$, $\alpha_{2}^{(1)}=0.5+0.5i$, $\alpha_{1}^{(2)}=0.45+0.45i$ and $\alpha_{2}^{(2)}=1+i$. }
\label{f11}
\end{figure*}
%%%%%%%%%%%%%%%%%%%%%%%%%%%%%%%%%%%%%%%%%%%%%%%
%%%%%%%%%%%%%%%%%%%%%%%%%%%%%%%%%%%%%%%%%%%%%%%
\begin{figure*}[t]
\centering
\includegraphics[width=0.95\linewidth]{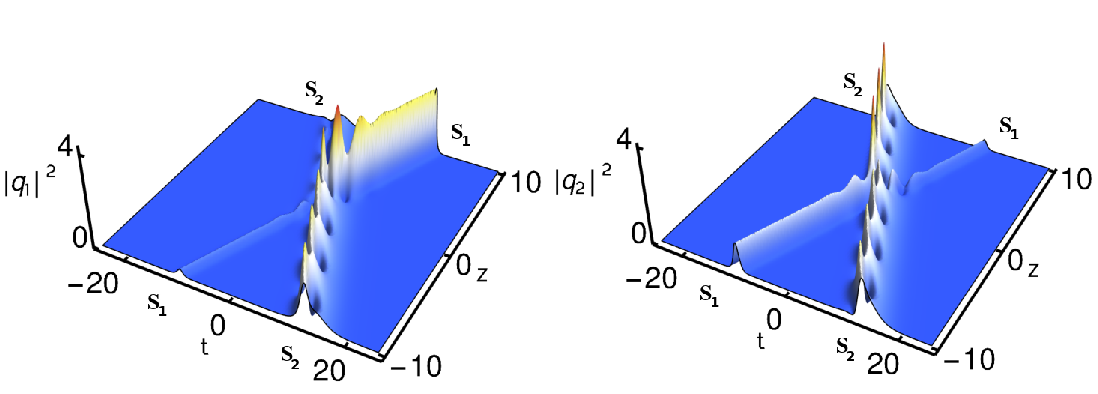}
\caption{Type-II energy sharing collision between the degenerate and nondegenerate solitons is illustrated by fixing the parameter values as $a=c=1$, $k_{1}=l_1=1.5+0.5i$, $k_{2}=1.5-0.5i$, $l_{2}=0.45-0.5i$, $\alpha_{2}^{(1)}=0.5$, $\alpha_{2}^{(2)}=0.45+0.45i$, $\alpha_{1}^{(1)}=0.5+0.5i$ and $\alpha_{1}^{(2)}=1+i$. }
\label{f12}
\end{figure*}
%%%%%%%%%%%%%%%%%%%%%%%%%%%%%%%%%%%%%%%%%%%
\section{Interaction between degenerate and nondegenerate solitons}
As we have pointed out earlier in Section II, the present GCNLS system (\ref{1}) can also admit degenerate and nondegenerate vector solitons simultaneously. Due to their coexistence it is of natural interest to investigate their collision dynamics. We find that they undergo the following two interesting types (Type-I and Type-II) of energy sharing collisions. As far as the Type-I energy sharing collision is concerned, both the degenerate as well as the nondegenerate solitons experience the same kind of energy sharing effect in all the modes. That is the degenerate soliton gets suppressed in its intensity in all the modes whereas the nondegenerate soliton gets enhanced in its intensity (or vice versa). On the contrary, in the Type-II energy sharing collision, the degenerate soliton undergoes opposite kind of energy switching collision with respect to the nondegenerate soliton. In this case, if the energy of the degenerate soliton is enhanced in one component (say $q_1$) its energy gets suppressed in the other component (say $q_2$). In this situation, the nondegenerate soliton exhibits opposite kind of energy switching collision in order to preserve the energy conservation. To investigate these two interesting collision scenarios, we again  analyze their analytical forms in the asymptotic limits $z\rightarrow\pm\infty$. In the following, we perform an asymptotic analysis for the first type of collision only. However, in principle, one can also carryout the calculations for the other case too in a similar manner.

\subsection{Asymptotic analysis: Type-I energy sharing collision}
In order to investigate the Type-I shape changing collision through the asymptotic analysis, 
we consider the parametric choice as follows: $k_{jR}$, $l_{2R}>0$, $j=1,2$, $k_{2I},l_{2I}>k_{1I}$, $k_{2I},l_{2I}>0$, and $k_{1I}<0$. This choice corresponds to a head-on collision between the degenerate and nondegenerate solitons. Using the above choice, we have to incorporate the asymptotic behavior of the wave variables, $\eta_{1R}=k_{1R}(t-2k_{1I}z)$, $\eta_{2R}=k_{2R}(t-2k_{2I}z)$, and   $\xi_{2R}=l_{2R}(t-2l_{2I}z)$ in the partially nondegenerate soliton solution (Eq. (\ref{3.7})  along with Eq. (\ref{3.10})) and deduce the asymptotic forms corresponding to the degenerate and nondegenerate solitons. The asymptotic behavior of the wave variables are found to be (i) Degenerate soliton 1 ($S_1$): $\eta_{1R}\simeq 0$, $\eta_{2R}$, $\xi_{2R}\rightarrow\pm \infty$ as $z\mp\infty$ and (ii) Nondegenerate soliton 2 ($S_2$): $\eta_{2R}$, $\xi_{2R}\simeq 0$, $\eta_{1R}\rightarrow\mp \infty$ as $z\mp\infty$. Under these asymptotic characters of wave variables, we deduce the following analytical forms of degenerate and nondegenerate solitons.\\ 
%\begin{widetext}\vspace{0.21cm}
(a) Before collision: $z\rightarrow -\infty$\\
 {\bf Degenerate soliton}: $\eta_{1R}\approx 0$, $\eta_{2R},~ \xi_{2R}\rightarrow +\infty$\\
In this limit, we deduce the corresponding asymptotic form of the degenerate soliton (say soliton 1) as
 \begin{eqnarray}
 	q_j\simeq A_j^- k_{1R}e^{i\eta_{1I}}\sech(\eta_{1R}+\frac{\hat{\lam}_5-\lam_{36}}{2}), ~j=1,2,
 	\label{20}
 \end{eqnarray}
 where $A_j^-=\frac{1}{(k_1+k_1^*)}e^{\Del_{1j}-\frac{\hat{\lam}_5+\lam_{36}}{2}}$. Here, the subscript $j$ denotes the modes and superscript $-$ represents the soliton before collision.  Again the various phase constants $\hat{\lambda}_5$, and $\lam_{36}$ are defined in Appendix C.\\
\\
{\bf Nondegenerate soliton}: $\eta_{2R},~\xi_{2R}\approx 0$, $\eta_{1R}\rightarrow -\infty$\\
 The following asymptotic expressions are deduced for the nondegenerate soliton (say soliton 2) and they read as
 {\scriptsize
 \bes
 \begin{eqnarray}
 	&&\hspace{-1.0cm}q_1=\frac{1}{D^-}\bigg(e^{i\eta_{2I}}e^{\frac{\ga_{20}+\rho_1'}{2}}\cosh(\xi_{2R}+\frac{\ga_{20}-\rho_1'}{2})+\frac{1}{2}e^{\ga_{15}}e^{i\xi_{2I}}[\cosh\eta_{2R}+\sinh\eta_{2R}]\bigg),\label{21a}\\
 	&&\hspace{-1.0cm}q_2=\frac{1}{D^-}\bigg(e^{i\xi_{2I}}e^{\frac{\ga_{15}+\rho_2'}{2}}\cosh(\eta_{2R}+\frac{\ga_{15}-\rho_2'}{2})+\frac{1}{2}e^{\nu_{20}}e^{i\eta_{2I}}[\cosh\xi_{2R}+\sinh\xi_{2R}]\bigg),\label{21b}\\
 	&&\hspace{-1.0cm}D^-=e^{\frac{\lam_{36}}{2}}\cosh(\eta_{2R}+\xi_{2R}+\frac{\lam_{36}}{2})+e^{\frac{\del_{4}+\del_{16}}{2}}\cosh(\eta_{2R}-\xi_{2R}+\frac{\del_{4}-\del_{16}}{2})\nonumber\\
 	&&\hspace{0.0cm}+e^{\frac{\del_{11}+\del_{12}}{2}}[\cosh(\frac{\del_{11}-\del_{12}}{2})\cos(\eta_{2I}-\xi_{2I})+i\sinh(\frac{\del_{11}-\del_{12}}{2})\sin(\eta_{2I}-\xi_{2I})],\label{21c}
 \end{eqnarray}
 \ees
 }
 where $e^{\rho_j'}=\al_2^{(j)}$, $j=1,2$.\\ 
(b) After collision: $z\rightarrow +\infty$\\
{\bf Degenerate Soliton}: $\eta_{1R}\approx0$,
 $\eta_{2R},~\xi_{2R}\rightarrow -\infty$\\
 The asymptotic form of the degenerate soliton is deduced from the partially nondegenerate soliton solution as follows:
	\begin{eqnarray}
		q_j\simeq A_j^+ k_{1R}e^{i\eta_{1I}}\sech(\eta_{1R}+\frac{R}{2}), ~j=1,2,
		\label{}
	\end{eqnarray}
where $\frac{R}{2}=\frac{1}{2}\log\frac{\Delta}{(k_1+k_1^*)^2}$, $A_j^+=\frac{\alpha_1^{(j)}}{e^{R/2}(k_1+k_1^*)}$, $j=1,2$, $\Delta=[a|\alpha_1^{(1)}|^2+c|\alpha_1^{(2)}|^2+b\alpha_1^{(1)}\alpha_1^{(2)*}+b^*\alpha_1^{(1)*}\alpha_1^{(2)}]$ . Here, $+$ denotes the soliton after collision.\\
{\bf Nondegenerate Soliton}: $\eta_{2R},~\xi_{2R}\approx 0$, $\eta_{1R}\rightarrow +\infty$,\\
In this limit, we deduced the form corresponding to the nondegenerate soliton after collision as 
 {\scriptsize
\bes
\begin{eqnarray}
	&&\hspace{-1.0cm}q_1=\frac{1}{D^+}\bigg(e^{i\xi_{2I}}e^{\frac{\hat{\mu}_2+\hat{\ga}_2}{2}}\cosh(\eta_{2R}+\frac{\hat{\mu}_2-\hat{\ga}_2}{2})+e^{i\eta_{2I}}e^{\frac{\hat{\mu}_1+\hat{\ga}_1}{2}}\cosh(\xi_{2R}+\frac{\hat{\mu}_1-\hat{\ga}_1}{2})\bigg),\label{23a}\\
	&&\hspace{-1.0cm}q_2=\frac{1}{D^+}\bigg(e^{i\xi_{2I}}e^{\frac{\hat{\chi}_2+\hat{\nu}_2}{2}}\cosh(\eta_{2R}+\frac{\hat{\chi}_2-\hat{\nu}_2}{2})+e^{i\eta_{2I}}e^{\frac{\hat{\chi}_1+\hat{\nu}_1}{2}}\cosh(\xi_{2R}+\frac{\hat{\chi}_1-\hat{\nu}_1}{2})\bigg),\label{23b}\\
	&&\hspace{-1.0cm}D^+=e^{\frac{\hat{\lam}_5+\Del_1}{2}}\cosh(\eta_{2R}+\xi_{2R}+\frac{\hat{\lam}_5-\Del_1}{2})+e^{\frac{\hat{\lam}_4+\hat{\lam}_1}{2}}\cosh(\eta_{2R}-\xi_{2R}+\frac{\hat{\lam}_1-\hat{\lam}_4}{2})\nonumber\\
	&&\hspace{0.0cm}+e^{\frac{\hat{\lam}_2+\hat{\lam}_3}{2}}[\cosh(\frac{\hat{\lam}_2-\hat{\lam}_3}{2})\cos(\eta_{2I}-\xi_{2I})+i\sinh(\frac{\hat{\lam}_2-\hat{\lam}_3}{2})\sin(\eta_{2I}-\xi_{2I})].\label{23c}
\end{eqnarray}
\ees
}
%\bes
%\bea
%&&q_1=\frac{1}{D^-}\big(e^{\eta_2+\hat{\ga}_1}+e^{\xi_2+\hat{\ga}_4}+e^{\eta_2+\eta_2^*+\xi_2+\hat{\mu}_{2}}+e^{\eta_2+\xi_2+\xi_2^*+\hat{\mu}_{1}}\big),\\
%&&q_2=\frac{1}{D^-}\big(e^{\eta_2+\hat{\nu}_1}+e^{\xi_2+\hat{\nu}_2}+e^{\eta_2+\eta_2^*+\xi_2+\hat{\chi}_{2}}+e^{\eta_2+\xi_2+\xi_2^*+\hat{\chi}_{1}}\big),
%\\
%&&D^-=e^{\Del}+e^{\eta_2+\eta_2^*+\hat{\lam}_1}+e^{\eta_2+\xi_2^*+\hat{\lam}_2}+e^{\eta_2^*+\xi_2+\hat{\lam}_3}+e^{\xi_2+\xi_2^*+\hat{\lam}_4}+e^{\eta_2+\eta_2^*+\xi_2+\xi_2^*+\hat{\lam}_5}.\nonumber
%\eea
%\ees
%\bes
%\bea
%&&q_1=\frac{1}{D^+}\big(\alpha_2^{(1)}e^{\eta_2}+e^{\eta_2+\eta_2^*+\xi_2+\ga_{15}}+e^{\eta_2+\xi_2+\xi_2^*+\ga_{20}}\big),\\
%&&q_2=\frac{1}{D^+}\big(\alpha_2^{(2)}e^{\xi_2}+e^{\eta_2+\eta_2^*+\xi_2+\nu_{15}}+e^{\eta_2+\xi_2+\xi_2^*+\nu_{20}}\big),\\
%&&D^+=1+e^{\eta_2+\eta_2^*+\del_4}+e^{\eta_2+\xi_2^*+\del_{11}}+e^{\eta_2^*+\xi_2+\del_{12}}+e^{\xi_2+\xi_2^*+\del_{16}}+e^{\eta_2+\eta_2^*+\xi_2+\xi_2^*+\lam_{36}},\nonumber
%\eea
%\ees
%\end{widetext}
We wish to note that the constants that are appearing in the above expressions are defined in Appendix C. 
%%%%%%%%%%%%%%%%%%%%%%%%%%%%%%%%%%%%%%%%%%%%%%%
\begin{figure*}[t]
\centering
\includegraphics[width=0.9\linewidth]{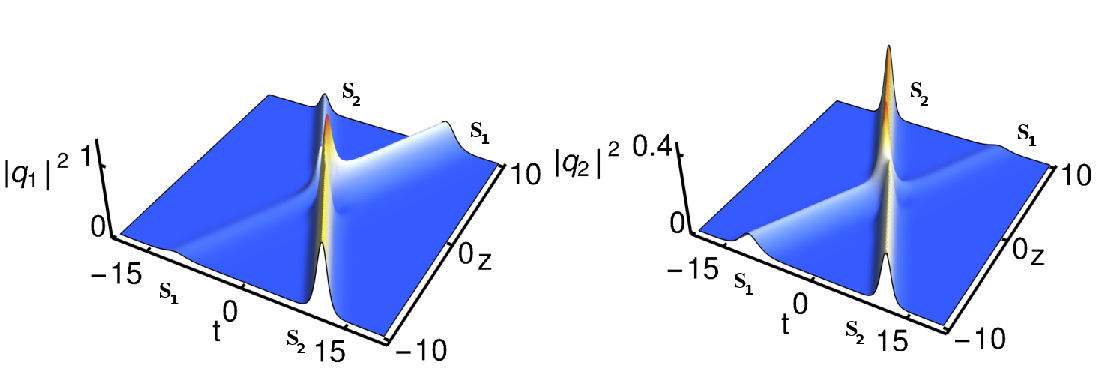}
\caption{Manakov type energy sharing collision among the two pure  degenerate solitons is illustrated by fixing the parameter values as $k_{1}=l_1=1-0.5i$, $k_{2}=l_2=0.5+0.5i$, $\alpha_{1}^{(1)}=0.5-0.5i$, $\alpha_{2}^{(1)}=0.5+0.5i$, $\alpha_{1}^{(2)}=0.8+0.25i$ and $\alpha_{2}^{(2)}=0.9$. }
\label{f13}
\end{figure*}
%%%%%%%%%%%%%%%%%%%%%%%%%%%%%%%%%%%%%%%
%%%%%%%%%%%%%%%%%%%%%%%%%%%%%%%%%%%%%%%%%%%%%%%
\begin{figure*}[t]
\centering
\includegraphics[width=0.9\linewidth]{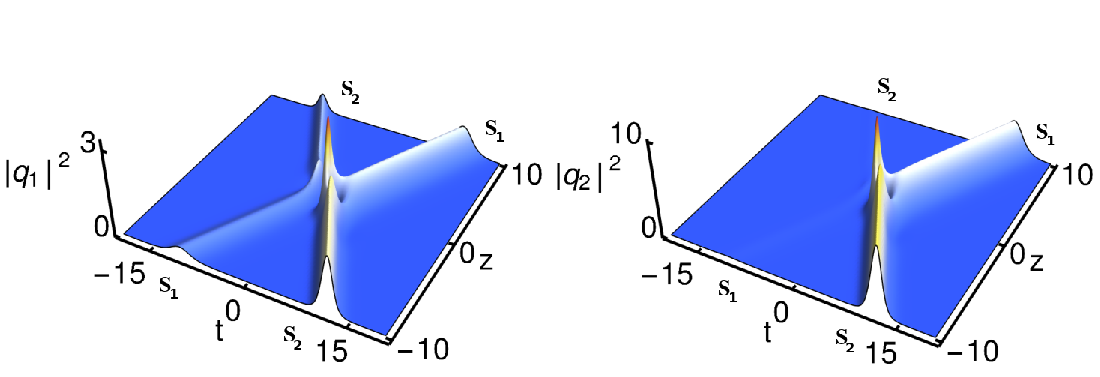}
\caption{Mixed CNLS case type energy sharing collision among the two pure  degenerate solitons is illustrated by fixing the parameter values as $k_{1}=l_1=1-0.5i$, $k_{2}=l_2=0.5+0.5i$, $\alpha_{1}^{(1)}=1.5-1.75i$, $\alpha_{2}^{(1)}=1.5$, $\alpha_{1}^{(2)}=0.5-0.5i$ and $\alpha_{2}^{(2)}=0.75i$.}
\label{f14}
\end{figure*}
%%%%%%%%%%%%%%%%%%%%%%%%%%%%%%%%%%%%%%%
%%%%%%%%%%%%%%%%%%%%%%%%%%%%%%%%%%%%%%%%%%%%%%%
\begin{figure*}[t]
\centering
\includegraphics[width=0.9\linewidth]{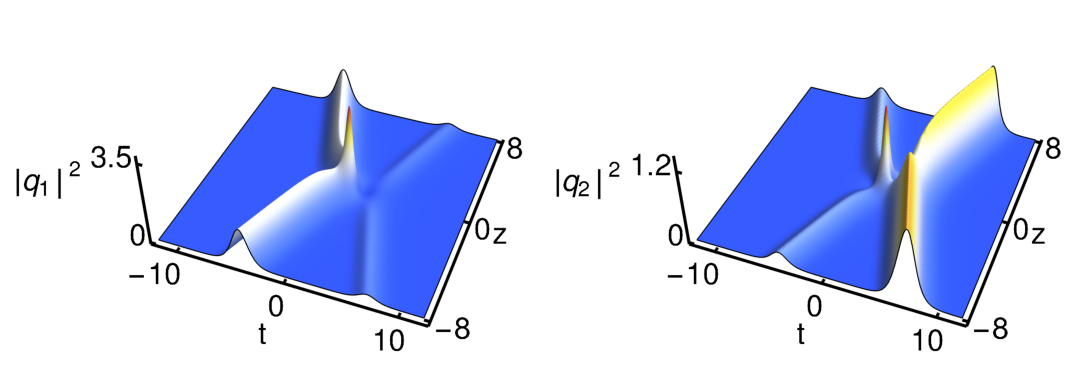}
\caption{Soliton reflection among the two pure  degenerate solitons is illustrated by fixing the parameter values as $k_{1}=l_1=1-0.5i$, $k_{2}=l_2=0.5+0.5i$, $\alpha_{1}^{(1)}=1.5-1.75i$, $\alpha_{2}^{(1)}=1.5$, $\alpha_{1}^{(2)}=0.5-0.5i$ and $\alpha_{2}^{(2)}=0.75i$. }
\label{f15}
\end{figure*}
%%%%%%%%%%%%%%%%%%%%%%%%%%%%%%%%%%%%%%%
\subsection{Energy sharing collisions between the degenerate and nondegenerate solitons}
As it is evident from the above asymptotic analysis, in Type-I energy sharing collision, both the degenerate soliton as well as the nondegenerate soliton experience shape changing nature during the collision process both in the cases of strong and weak FWM effects. As far as the degenerate soliton is concerned, the amplitude of it changes from $A_j^-k_{1R}$ (before collision) to $A_j^+k_{1R}$ (after collision). Then, for the nondegenerate soliton, the asymptotic expressions as well as the phase terms do not preserve their forms and they are drastically varied during the collision process.  This implies that there is a definite possibility of observing shape changing collision between the degenerate soliton and the nondegenerate soliton. However, the mechanism behind the shape changing behavior of the degenerate soliton is distinct from the nondegenerate soliton. The degenerate soliton, as we have expected, undergoes shape changing behavior by sharing its energy to the nondegenerate soliton. In this case, the polarization vectors, $A_j^{\pm}$, of the degenerate soliton play dominant roles for the energy redistribution of the nondegenerate soliton in all the modes. In contrast to this, in the case of nondegenerate soliton, the relative separation distances (or phase terms)  do not remain constant throughout the collision process and it gains energy from the degenerate soliton. A typical energy sharing collision of the first type is demonstrated in Fig. \ref{f11}, where the intensity of the breathing nondegenerate soliton ($S_2$) (or degenerate soliton ($S_1$)) is enhanced (or suppressed) in both the modes along with a finite phase shift. In this collision scenario, the nondegenerate soliton gains energy from the degenerate soliton. Such energy redistribution can be characterized by calculating the transition amplitude of the degenerate soliton. The transition amplitude of the degenerate soliton is calculated from its corresponding asymptotic expressions before and after collision as
\begin{equation}
	T_j^1=\frac{A_j^+}{A_j^-}=\frac{\al_1^{(j)}e^{\frac{\hat{\lam}_5+\lam_{36}-R}{2}}}{e^{\Del_{1j}}}, ~j=1,2.\label{24}
\end{equation}
Here, the subscript $j$ represents $j$th mode and the superscript 1 denotes the soliton 1 (or degenerate soliton). One can also calculate the change in the intensity of the degenerate soliton by simply taking the absolute square of the transition amplitudes $T_j^l$. That is,  
\begin{equation}
	|T_j^1|^2=\frac{|A_j^+|^2}{|A_j^-|^2}=\frac{|\al_1^{(j)}e^{\frac{\hat{\lam}_5+\lam_{36}}{2}}|^2}{|e^{\Del_{1j}+\frac{R}{2}}|^2}, ~j=1,2. \label{25}
\end{equation} 
The variations of the phase terms, in the nondegenerate soliton case,  can be calculated from the expressions (\ref{21a})-(\ref{21c})  and (\ref{23a})-(\ref{23c}). For brevity, we have omitted the details. Further, we find that the periodic nature of the nondegenerate soliton is preserved throughout the collision process and subsequently the time period of oscillation, $T=\frac{2\pi}{k_{2R}^2-l_{2R}^2}$, remains constant. However, in Type-I energy sharing collision, as per Eqs. (\ref{17}) and (\ref{18}), the total energy of the individual solitons in both the modes $q_1$ and $q_2$ are conserved and also the intensity of the individual modes are conserved.  

As we mentioned earlier, we also observe another interesting energy sharing collision between the degenerate soliton ($S_1$) and breathing nondegenerate soliton ($S_2$). Such a collision scenario is depicted in Fig. \ref{f12}, from which one can observe that the energy of the degenerate soliton gets enhanced in the first mode whereas it gets suppressed in the other mode. To hold the energy conservation (through Eq. \ref{17}) in the individual mode, the nondegenerate soliton undergoes opposite kind of energy switching collision. That is the intensity of oscillations of the nondegenerate soliton gets suppressed in the first mode while it gets enhanced in the second mode. In this case also, the periodic nature of the nondegenerate soliton does not change under collision with the degenerate one. To analyze the energy redistribution nature of this collision further one can perform the asymptotic analysis as we have done earlier for the case of Type-I energy sharing collision. By doing so, one would find the transition amplitudes associated with the degenerate soliton and the variation in relative separation distance of the nondegenerate soliton. We wish to point out that both Type-I and Type-II energy sharing collisions presented here have not been observed earlier in the literature in the GCNLS system (\ref{1}). We also wish to point out that the type-II energy sharing collision is quite similar to the collision scenario among the degenerate Manakov solitons \cite{radha}. However, the mechanism behind each of them is entirely different. The shape changing properties both degenerate and nondegenerate solitons are useful for manipulating light by light through their collision.
%\clearpage   \twocolumngrid
%%%%%%%%%%%%%%%%%%%%%%%%%%%%%%%%%%%%%%%%%%%%%%%
\section{Collision dynamics among the pure degenerate solitons}
Now, for completeness, we wish to indicate the interactions among the two completely degenerate solitons. To bring out the corresponding collision scenario one has to consider either the nondegenerate two-soliton solution (\ref{3.7}), with $k_1=l_1$, $k_2=l_2$, or Eq. (\ref{3.7}) along with Eq. (\ref{12}). Such wave number restrictions and suitable choice of complex phase constants $\al_{1,2}^{(j)}$, $j=1,2$, yield interesting shape changing collisions. It is well known that the degenerate solitons in the present GCNLS system (\ref{1}) exhibit three kinds of shape changing or energy sharing collisions for three different choices of SPM ($a$), XPM ($c$), FWM ($b$)  nonlinearities. They are referred as follows: (i) Manakov type shape changing collision: $a=c=1$, $b\neq 0$, (ii) Mixed CNLS type shape changing collision:  $a=-c=1$, $b\neq 0$, and (iii) Soliton reflection like shape changing collision: $a=c=0$, $b\neq 0$. The degenerate solitons share energy among themselves by following energy or intensity redistribution mechanism. A typical Manakov type energy sharing collision is demonstrated in Fig. \ref{f13} for $a=c=b=1$, and the other parameters are given in the corresponding figure caption. From the latter figure, one can observe that the degenerate soliton $S_2$ undergoes intensity suppression in the first mode and it gets enhanced in the second mode and the reverse collision scenario take place in the other degenerate soliton $S_1$ in order to hold the energy conservation. In this case, the total energy of the degenerate solitons in both the modes is conserved and the energy conservation in individual modes is also preserved. This kind of intensity redistribution comes out because of the variation in the polarizations of the degenerate solitons. 

 The present GCNLS system (\ref{1}) also admits another interesting collision scenario, which is quite similar to the one observed as in the case of mixed CNLS system \cite{mix}. Such a shape changing collision is demonstrated in Fig. \ref{f14} for $a=-c=1$, $b=1$. The figure shows that the given degenerate soliton (say $S_1$)  exhibits the same type of energy change in both the modes. For instance, in Fig. \ref{f14}, the energy of soliton 1 gets enhanced in both the modes whereas the intensity of soliton 2 gets suppressed in all the modes. In addition, the degenerate solitons undergo a third type of energy sharing collision as it is illustrated in Fig. \ref{f15}, for $a=c=0$, $b=1$. During this collision scenario the two solitons undergo an interaction which is quite similar to the Manakov type shape changing collision (Fig. \ref{f13}). However, from Fig. \ref{f15}, one can observe that the two degenerate solitons come close together and they are bounced back by the collision. After the collision process, they stay away from each other with a finite change in their intensities. This kind collision behavior is referred as soliton reflection in Ref. \cite{yang1}. In fact the soliton reflection demonstrated in Fig. \ref{f15} is quite distinct from the one that was pointed out in Ref. \cite{yang1}, where the first soliton in all the modes has higher power than the second one. We wish to point out that in all the three cases the degenerate solitons experience amplitude dependent phase shifts which leads to appropriate change in the relative separation distance between the solitons before and after collision.

\section{Conclusion}
In this paper, to investigate the effect of four-wave mixing phenomenon on the structure and collision dynamics of nondegenerate vector solitons, we have considered a generalized coupled nonlinear Schr\"{o}dinger system. The fundamental and higher-order nondegenerate vector soliton solutions, including the general $N$-soliton solution, are obtained through the Hirota bilinear method and their forms are rewritten in a compact way using Gram determinants. We found that the presence of FWM  induces a breathing vector soliton state in both the optical modes. Such breather formation is not possible in the fundamental degenerate vector bright solitons of the present GCNLS system (\ref{1}) as well as in the fundamental vector solitons (both degenerate and nondegenerate cases) of the Manakov and mixed CNLS systems. Then, we have observed in the present GCNLS system the nondegenerate solitons, in general, undergo a novel shape changing collisions for both strong and weak FWM effects. However, under an appropriate choice of propagation constants, they also exhibit a shape preserving collision.  Further, by imposing a restriction on the wave wave numbers we have deduced the partially nondegenerate two-soliton solution from the completely nondegenerate two-soliton solution. The existence of such interesting class of two-soliton solution immediately gave us freedom to analyze the interaction between the degenerate and nondegenerate solitons. While analyzing the collision between them we found that they undergo two types of energy sharing collisions. In each of these collision scenarios, the shape changing nature happened in the degenerate soliton due to its polarization variation whereas in the nondegenerate case its due to a drastic alteration in phases or relative separation distance.   To the best of our knowledge the latter collision scenarios as well as the collision scenarios among the two nondegenerate solitons have not been reported earlier in the literature. For completeness, the various energy sharing collision scenarios related to the pure degenerate bright solitons are indicated. We believe that the results reported in this paper will be useful in nonlinear optics for manipulating light by light through collision. In principle, this study can also be extended to other physically important integrable systems, where the dynamics of optical modes are governed by the coupled nonlinear Schr\"odinger models. It is also interesting to study the nondegenerate vector solitons in dissipative systems.\\

%\begin{acknowledgements}
\hspace*{-0.5cm}{\bf Acknowledgements}
The works of M. Kirane, and S. Stalin, are supported by Khalifa University of Science and Technology, Abu-Dhabi, UAE, under the Project Grant No. 8474000355. M. Lakshmanan thanks DST-SERB, INDIA for the award of a DST-SERB National Science Chair (NSC/2020/000029) position in which R. Ramakrishnan is currently working as a Research Associate. R. Ramakrishnan is also acknowledges CSIR, INDIA for the financial support in the form of Direct-Senior Research Fellowship (09/475(0203)/2020-EMR-I) during the period 2021-2022.\\
%\end{acknowledgements}

\hspace*{-0.5cm}{\bf Data availability}
The authors declare that all data generated or analyzed during this study are included in this article.\\

\hspace*{-0.5cm}{\bf Conflict of interest}
The authors declare that they have no conflict of interest concerning the publication of this manuscript.

%\appendix 
\section*{Appendix A: Nondegenerate N-soliton solution}\label{A}
By following the procedure described in Section 2, one can obtain the general form of nondegenerate N-soliton solution. The form turns out to be
\begin{eqnarray}
\hspace{-0.5cm}g^{(s)}&=&
\begin{vmatrix}
	A & I & \phi \\ 
	-I &  B & {\bf 0}^T\\
	{\bf 0} & C_s & 0
\end{vmatrix},~f=\begin{vmatrix}
		A & I \\ 
	-I &  B 
\end{vmatrix},~s=1,2,\label{3.A1}
\end{eqnarray}
where the various elements of matrices $A$ and $B$ are defined as \begin{eqnarray}	
A =\begin{pmatrix}
A_{mm'}& A_{mn} \\
A_{nm} & A_{nn'}
\end{pmatrix},~
B =\begin{pmatrix}
\kappa_{mm'}& \kappa_{mn} \\
\kappa_{nm} & \kappa_{nn'}	
\end{pmatrix},
\end{eqnarray}	
\begin{eqnarray}		
&&A_{mm'} = \frac{e^{\eta_m+\eta^*_{m'}}}{(k_m+k^*_{m'})}, 
~A_{mn} = \frac{e^{\eta_m+\xi^*_n}}{(k_m+l^*_n)},	
~A_{nm} = \frac{e^{\eta^*_n+\xi_m}}{(k^*_n+l_m)},
~A_{nn'} = \frac{e^{\xi_n+\xi^*_{n'}}}{(l_n+l^*_{n'})}, \nonumber \\
&&\kappa_{mm'}=\frac{\psi_m^{\dagger}\sigma\psi_{m'}}{(k_m^*+k_{m'})},
~\kappa_{mn}=\frac{\psi_m^{\dagger}\sigma\psi'_{n}}{(k_m^*+l_{n})},
~\kappa_{nm}=\frac{\psi_n^{'\dagger}\sigma\psi_{m}}{(l_n^*+k_{m})},
~\kappa_{nn'}=\frac{\psi_n^{'\dagger}\sigma\psi'_{n'}}{(l_n^*+l_{n'})}, \nonumber \\
&& m,m',n,n'=1,2,3.\nonumber
\end{eqnarray}
In (\ref{3.A1}) the column matrices are 
$\psi_j=\begin{pmatrix}
	\alpha_j^{(1)}\\
	0
\end{pmatrix}$, ~$\psi'_j=\begin{pmatrix}
	0\\
	\alpha_j^{(2)}
\end{pmatrix}$, $j=m,m',n,n'=1,2,...,N$, $\eta_j=k_jt+ik_j^2z$ and $\xi_j=l_jt+il_j^2z$, $j=1,2,..,N$.
The other matrices in Eq. (\ref{3.A1}) are defined below: \\
$\phi=\begin{pmatrix}
	e^{\eta_1} & e^{\eta_2}  &\cdot&\cdot &e^{ \eta_N} & e^{\xi_1}  & e^{\xi_2}  &\cdot &\cdot &e^{ \xi_N}
\end{pmatrix}^T$, $C_1=-\begin{pmatrix}
	\alpha_1^{(1)} & \alpha_2^{(1)} & \cdot & \cdot& \alpha_N^{(1)} & 0 &0& \cdot& \cdot&0
\end{pmatrix}$, $C_2=-\begin{pmatrix}
	0 & 0 & \cdot & \cdot & 0 &\alpha_1^{(2)} & \alpha_2^{(2)} & \cdot& \cdot& \alpha_N^{(2)} \end{pmatrix}$, ${\bf 0} =\begin{pmatrix}
	0 & 0  & \cdot & \cdot& 0 \end{pmatrix}$, $\sigma=\begin{pmatrix}
	a&b^*\\
	b&c
\end{pmatrix}$, and  $I$ is a $(N\times N)$ identity matrix.
%\vspace{-10.0cm}
\section*{Appendix B: The various constants which appear in Section 3}\label{B}
By defining the various quantities,
\begin{eqnarray*}
\kappa_{ij}=\frac{1}{k_i^*+k_j},~\kappa_{21}=\kappa_{12}^*,~\theta_{ij}=\frac{1}{k_i+l_j^*},~n_{ij}=\frac{1}{l_i+l_j^*},
~n_{21}=n_{12}^*,~i,j=1,2,
\end{eqnarray*}
we can introduce the following constants:
\begin{eqnarray*}
&&e^{\ga_2}=cn_{11}(n_{11}-\theta_{11})\al_1^{(1)}|\al_1^{(2)}|^2,~e^{\ga_1}=b^*\theta_{11}^*(\theta_{11}^*-\kappa_{11})|\al_1^{(1)}|^2\al_1^{(2)},~e^{\del_1}=a|\al_1^{(1)}|^2\kappa_{11}^2,~e^{\del_5}=b\al_1^{(1)}\al_1^{(2)*}\theta_{11}^2,\\
&&e^{\del_6}=b^*\al_1^{(1)*}\al_1^{(2)}\theta_{11}^{*2},~e^{\del_{13}}=cn_{11}^2|\al_1^{(2)}|^2,~e^{\lam_1}=|\al_1^{(1)}|^2|\al_1^{(2)}|^2[|\theta_{11}|^2-n_{11}\kappa_{11}][|b|^2|\theta_{11}|^2-acn_{11}\kappa_{11}],\\
&&e^{\nu_1}=a\kappa_{11}(\kappa_{11}-\theta_{11}^*)\al_1^{(2)}|\al_1^{(1)}|^2,~e^{\nu_2}=b\theta_{11}(\theta_{11}-n_{11})\al_1^{(1)}|\al_1^{(2)}|^2,\\
&&e^{\mu_{1}}=\al_2^{(1)}|\al_1^{(1)}|^2|\al_1^{(2)}|^2\big(n_{11}\kappa_{12}^*+\theta_{11}(\theta_{11}^*-\kappa_{12}^*)+\theta_{21}\kappa_{11}-n_{11}\kappa_{11}-\theta_{11}^*\theta_{21}\big)\big(|b|^2\theta_{11}^*(\theta_{11}-\theta_{21})+acn_{11}(\kappa_{12}^*-\kappa_{11})\big),\\
%\end{eqnarray*}
%\begin{eqnarray*}
&&e^{\mu_4}=b^*c|\al_1^{(1)}|^2|\al_1^{(2)}|^2\al_2^{(2)}\big(n_{12}^*\theta_{11}^*-n_{11}\theta_{12}^*\big)\big(\theta_{11}(\theta_{12}^*-\theta_{11}^*)+n_{12}^*(\theta_{11}^*-\kappa_{11})+n_{11}(\kappa_{11}-\theta_{12}^*)\big),\\
&&e^{\mu_{28}}=c|\al_1^{(1)}|^2|\al_1^{(2)}|^2|\al_2^{(2)}|^2\al_2^{(1)}\big[|b|^2\big(n_{22}\theta_{11}^*(\theta_{11}-\theta_{21})+n_{12}^*\theta_{11}^*(\theta_{22}-\theta_{12})\\
&&\hspace{1.0cm}+\theta_{12}^*(-n_{12}\theta_{11}+n_{11}\theta_{12}+n_{12}\theta_{21}-n_{11}\theta_{22})\big)+ac(\kappa_{11}-\kappa_{12}^*)(|n_{12}|^2-n_{11}n_{22})\big]\\
&&\hspace{1.0cm}\bigg[-n_{12}\theta_{11}\theta_{12}^*+n_{11}|\theta_{12}|^2+\theta_{11}^*\theta_{12}\theta_{21}+n_{12}\theta_{12}^*\theta_{21}-|\theta_{12}|^2\theta_{21} 
-|\theta_{11}|^2\theta_{22}-n_{11}\theta_{12}^*\theta_{22}+\theta_{11}\theta_{12}^*\theta_{22}\\
&&\hspace{1.0cm}+(-n_{12}\theta_{21}+n_{11}\theta_{22})\kappa_{11}+(n_{12}\theta_{11}-n_{11}\theta_{12})\kappa_{12}^*+n_{22}\big(-\theta_{11}^*\theta_{21}-n_{11}\kappa_{11}+\theta_{21}\kappa_{11}+\theta_{11}(\theta_{11}^*-\kappa_{12}^*)\\
&&\hspace{1.0cm}+n_{11}\kappa_{12}^*\big)+n_{12}^*\big(-\theta_{11}^*\theta_{12}+\theta_{11}^*\theta_{22}+n_{12}\kappa_{11}-\theta_{22}\kappa_{11}-n_{12}\kappa_{12}^*+\theta_{12}\kappa_{12}^*\big)\bigg],\\
&&e^{\mu_{27}}=\theta_{11}^*\theta_{21}\theta_{22}^*-n_{12}^*\theta_{21}^*\kappa_{11}+|\theta_{21}|^2\kappa_{11}+n_{11}\theta_{22}^*\kappa_{11}-\theta_{21}\theta_{22}^*\kappa_{11}+\kappa_{12}(n_{12}^*\theta_{11}^*-\theta_{11}^*\theta_{21})\\
&&\hspace{1.0cm}+\kappa_{12}^*(n_{12}^*\theta_{21}^*-n_{11}\theta_{22}^*+n_{11}|\kappa_{12}|^2	
-n_{12}^*|\kappa_{12}|^2-n_{12}^*\theta_{11}^*\kappa_{22}+(-n_{11}+n_{12})\kappa_{11}\kappa_{22}\\
&&\hspace{1.0cm}+\theta_{12}^*(-|\theta_{21}|^2-n_{11}\kappa_{12}+\theta_{21}\kappa_{12}+n_{11}\kappa_{22})
+\theta_{11}(-\theta_{11}^*\theta_{22}^*-\theta_{21}^*\kappa_{12}^*+\theta_{22}^*\kappa_{12}^*+\theta_{12}^*(\theta_{21}^*-\kappa_{22})+\theta_{11}^*\kappa_{22})),\\	
&&e^{r_8}=c|\al_1^{(1)}|^2|\al_1^{(2)}|^2|\al_2^{(2)}|^2\big(n_{22}|\theta_{11}|^2-n_{12}^*\theta_{11}^*\theta_{12}-n_{12}\theta_{11}\theta_{12}^*+n_{11}|\theta_{12}|^2+\kappa_{11}|n_{12}|^2-n_{11}n_{22}\kappa_{11}\big)\\
&&\hspace{1.0cm}\times\bigg[|b|^2\big(n_{22}|\theta_{11}|^2-n_{12}^*\theta_{11}^*\theta_{12}-n_{12}\theta_{11}\theta_{12}^*+n_{11}|\theta_{12}|^2\big)+ac\kappa_{11}(|n_{12}|^2-n_{11}n_{22})\bigg],\\
&&e^{r_2}=b^*|\al_1^{(1)}|^2|\al_1^{(2)}|^2\al_2^{(1)*}\al_2^{(2)}\big(\theta_{11}\theta_{12}^*\theta_{21}^*-|\theta_{11}|^2\theta_{22}^*-n_{12}^*\theta_{21}^*\kappa_{11}+n_{11}\theta_{22}^*\kappa_{11}+(n_{12}^*\theta_{11}^*-n_{11}\theta_{12}^*)\kappa_{12}\big)\\
&&\hspace{1.0cm}\times\bigg[|b|^2\theta_{11}(\theta_{12}^*\theta_{21}^*-\theta_{11}^*\theta_{22}^*)+ac((n_{11}\theta_{22}^*-n_{12}^*\theta_{21}^*)\kappa_{11}+\kappa_{12}(n_{12}^*\theta_{11}^*-n_{11}\theta_{12}^*))\bigg],\\
\end{eqnarray*}
\begin{eqnarray*}
&&e^{r_5}=b|\al_1^{(1)}|^2|\al_1^{(2)}|^2\al_2^{(1)}\al_2^{(2)*}\big(\theta_{11}^*\theta_{12}\theta_{21}-|\theta_{11}|^2\theta_{22}-n_{12}\theta_{21}\kappa_{11}+n_{11}\theta_{22}\kappa_{11}+n_{12}\theta_{11}\kappa_{12}^*-n_{11}\theta_{12}\kappa_{12}^*\big)\\	
&&\hspace{1.0cm}\times\bigg[|b|^2\theta_{11}^*(\theta_{12}\theta_{21}-\theta_{11}\theta_{22})+ac((n_{11}\theta_{22}-n_{12}\theta_{21})\kappa_{11}+(n_{12}\theta_{11}-n_{11}\theta_{12})\kappa_{12}^*)\bigg],\\	
&&e^{r_1}=\big[|b|^2(|\theta_{21}|^2\kappa_{11}-\theta_{11}^*\theta_{21}\kappa_{12}-\theta_{11}\theta_{21}^*\kappa_{12}^*+|\theta_{11}|^2\kappa_{22})+acn_{11}(|\kappa_{12}|^2-\kappa_{11}\kappa_{22})\big],\\
&&e^{r_{17}}=\bigg[(n_{12}\theta_{21}-n_{11}\theta_{22})(\theta_{22}^*\kappa_{11}-\theta_{12}^*\kappa_{12})+n_{22}(-|\theta_{21}|^2\kappa_{11}+\theta_{11}^*\theta_{21}\kappa_{12}-n_{11}|\kappa_{12}|^2+n_{11}\kappa_{11}\kappa_{22})\\	
&&\hspace{1.0cm}n_{12}^*(\theta_{21}^*\theta_{22}\kappa_{11}-\theta_{11}^*\theta_{22}\kappa_{12}+n_{12}|\kappa_{12}|^2-n_{12}\kappa_{11}\kappa_{22})+\theta_{12}(\theta_{12}^*|\theta_{21}|^2-\theta_{11}^*\theta_{21}\theta_{22}^*-n_{12}^*\theta_{21}^*\kappa_{12}^*\\	
&&\hspace{1.0cm}+n_{11}\theta_{22}^*\kappa_{12}^*+n_{12}^*\theta_{11}^*\kappa_{22}-n_{11}\theta_{12}^*\kappa_{22})+\theta_{11}(\theta_{11}^*|\theta_{22}|^2-\theta_{12}^*\theta_{21}^*\theta_{22}+n_{22}\theta_{21}^*\kappa_{12}^*-n_{12}\theta_{22}^*\kappa_{12}^*\\
&&\hspace{1.0cm}-n_{22}\theta_{11}^*\kappa_{22}+n_{12}\theta_{12}^*\kappa_{22})\bigg]	
\times\bigg[|b|^4(\theta_{12}\theta_{21}-\theta_{11}\theta_{22})(\theta_{12}^*\theta_{21}^*-\theta_{11}^*\theta_{22}^*)+a^2c^2(|n_{12}|^2-n_{11}n_{22})\\
&&\hspace{1.0cm}(|\kappa_{12}|^2-\kappa_{11}\kappa_{22})+ac|b|^2\big((n_{12}\theta_{21}\theta_{22}-n_{11}|\theta_{22}|^2)\kappa_{11}
+n_{11}(\theta_{12}^*\theta_{22}\kappa_{12}+\theta_{12}\theta_{22}^*\kappa_{12}^*)\\
&&\hspace{1.0cm}-n_{12}(\theta_{12}^*\theta_{21}\kappa_{12}+\theta_{11}\theta_{22}^*\kappa_{12}^*)+(n_{12}\theta_{11}\theta_{12}^*-n_{11}|\theta_{12}|^2)\kappa_{22}+n_{22}(\theta_{11}^*\theta_{21}\kappa_{12}-|\theta_{21}|^2\kappa_{11}\\
&&\hspace{1.0cm}+\theta_{11}\theta_{21}^*\kappa_{12}^*-|\theta_{11}|^2\kappa_{22})+n_{12}^*(\theta_{21}^*\theta_{22}\kappa_{11}-\theta_{11}^*\theta_{22}\kappa_{12}-\theta_{12}\theta_{21}^*\kappa_{12}^*+\theta_{11}^*\theta_{12}\kappa_{22})\big)\bigg],\\
&&e^{\chi_4}=|\al_1^{(1)}|^2|\al_1^{(2)}|^2\al_2^{(2)}\big[|b|^2\theta_{11}(\theta_{11}^*-\theta_{12}^*)+ac\kappa_{11}(n_{11}-n_{12}^*)\big]\big[\theta_{11}(\theta_{12}^*-\theta_{11}^*)+n_{12}^*(\theta_{11}^*-\kappa_{11})+n_{11}(\kappa_{11}-\theta_{12}^*)\big],\\
&&e^{\chi_1}=ab|\al_1^{(1)}|^2|\al_1^{(2)}|^2\al_2^{(1)}[\theta_{21}\kappa_{11}-\theta_{11}\kappa_{12}^*][-\theta_{11}^*\theta_{21}-n_{11}\kappa_{11}+\theta_{21}\kappa_{11}+\theta_{11}(\theta_{11}^*-\kappa_{12}^*)+n_{11}\kappa_{12}^*]\\	
&&\hspace{1.0cm}-b|\al_1^{(1)}|^2|\al_1^{(2)}|^2|\al_2^{(2)}|^2\al_2^{(1)}[-|b|^2(\theta_{11}^*-\theta_{12}^*)(\theta_{12}\theta_{21}-\theta_{11}\theta_{22})+ac\big((n_{12}\theta_{21}-n_{22}\theta_{21}+\theta_{12}^*\theta_{22})\kappa_{11}\\	
&&\hspace{1.0cm}+\kappa_{12}^*(n_{22}-n_{12})\theta_{11}+(n_{11}+n_{12}^*)\theta_{12}\kappa_{12}^*\big)],\\
%\end{eqnarray*}
%\begin{eqnarray*}
&&e^{\chi_{26}}=-a|\al_1^{(1)}|^2|\al_1^{(2)}|^2|\al_2^{(1)}|^2\al_2^{(2)}\big[-ac(n_{11}-n_{12}^*)(|\kappa_{12}|^2-\kappa_{11}\kappa_{22})+|b|^2\big(\theta_{21}(-\theta_{21}^*\kappa_{11}+\theta_{22}^*\kappa_{11}+\kappa_{12}(\theta_{11}^*-\theta_{12}^*))\\	
&&\hspace{1.0cm}+\theta_{11}(\kappa_{12}^*(\theta_{21}^*-\theta_{22}^*)+\kappa_{22}(\theta_{12}^*-\theta_{11}^*))\big)\big]\bigg[\theta_{11}^*\theta_{21}\theta_{22}^*-n_{12}^*\theta_{21}^*\kappa_{11}+|\theta_{21}|^2\kappa_{11}+n_{11}\theta_{22}^*\kappa_{11}-\theta_{21}\theta_{22}^*\kappa_{11}\\ &&\hspace{1.0cm}+(n_{12}^*\theta_{11}^*-\theta_{11}^*\theta_{21})\kappa_{12}+(n_{12}^*\theta_{21}^*-n_{11}\theta_{22}^*)\kappa_{12}^*-n_{11}|\kappa_{12}|^2-n_{12}^*|\kappa_{12}|^2-n_{12}^*\theta_{11}^*\kappa_{22}-n_{11}\kappa_{11}\kappa_{22}+n_{12}^*\kappa_{11}\kappa_{22}\\	
&&\hspace{1.0cm}+\theta_{12}^*(-|\theta_{21}|^2-n_{11}\kappa_{12}+\theta_{21}\kappa_{12}+n_{11}\kappa_{22})+\theta_{11}(-\theta_{11}^*-\theta_{22}^*-\theta_{21}^*\kappa_{12}^*+\theta_{22}^*\kappa_{12}^*+\theta_{12}^*(\theta_{21}^*-\kappa_{22})+\theta_{11}^*\kappa_{22})\bigg],\\
&&e^{\chi_{27}}=-b|\al_1^{(1)}|^2|\al_1^{(2)}|^2|\al_2^{(2)}|^2\al_2^{(1)}\big[-|b|^2(\theta_{11}^*-\theta_{12}^*)(\theta_{12}\theta_{21}-\theta_{11}\theta_{22})+ac\big((n_{12}-n_{22})\theta_{21}\kappa_{11}+\theta_{12}^*\theta_{22}\kappa_{11}\\	
&&\hspace{1.0cm}+\kappa_{12}^*(n_{22}\theta_{11}-n_{12}\theta_{11})+\kappa_{12}^*(n_{11}\theta_{12}-n_{12}^*\theta_{12})\big)\big]\bigg[n_{11}|\theta_{12}|^2-n_{12}\theta_{11}\theta_{12}^*+(\theta_{11}^*\theta_{12}+n_{12}\theta_{12}^*)\theta_{21}-|\theta_{12}|^2\theta_{21}\\	
&&\hspace{1.0cm}-|\theta_{11}|^2\theta_{22}+(\theta_{11}-n_{11})\theta_{12}^*\theta_{22}+(n_{11}\theta_{22}-n_{12}\theta_{21})\kappa_{11}+(n_{12}\theta_{11}-n_{11}\theta_{12})\kappa_{12}^*+n_{22}(\theta_{11}^*\theta_{21}-n_{11}\kappa_{11}\\	
&&\hspace{1.0cm}+\theta_{21}\kappa_{11}+\theta_{11}(\theta_{11}^*-\kappa_{12}^*)+n_{11}\kappa_{12}^*)+n_{12}^*(\theta_{11}^*\theta_{22}-\theta_{11}^*\theta_{12}+(n_{12}-\theta_{22})\kappa_{11}+(\theta_{12}-n_{12})\kappa_{12}^*)\bigg],\\
%\end{eqnarray*}
%\begin{eqnarray*}
&&e^{\mu_{24}}=b^*c\al_1^{(2)}|\al_2^{(1)}|^2|\al_2^{(2)}|^2(n_{22}\theta_{21}^*-n_{12}\theta_{22}^*)\big[-\theta_{21}^*\theta_{22}-n_{12}\theta_{22}^*+|\theta_{22}|^2+n_{12}\kappa_{22}+n_{22}(\theta_{21}^*-\kappa_{22})\big],\\
&&e^{\mu_{23}}=\al_1^{(1)}|\al_2^{(1)}|^2|\al_2^{(2)}|^2\big[\theta_{22}(\kappa_{12}-\theta_{22}^*)+\theta_{12}(\theta_{22}^*-\kappa_{22})+n_{12}(\kappa_{22}-\kappa_{12})\big]\big(|b|^2\theta_{22}^*(\theta_{12}-\theta_{22})+acn_{22}(\kappa_{22}-\kappa_{12})\big),\\	
&&e^{\mu_{26}}=c|\al_1^{(2)}|^2|\al_2^{(1)}|^2|\al_2^{(2)}|^2\big[|b|^2\big(n_{22}\theta_{21}^*(\theta_{11}-\theta_{21})+n_{12}^*\theta_{21}^*(\theta_{22}-\theta_{21})+\theta_{22}^*(-n_{12}\theta_{11}+n_{11}\theta_{12}+n_{12}\theta_{21}-n_{11}\theta_{22})\big)\\
&&\hspace{1.0cm}+ac(\kappa_{12}-\kappa_{22})(|n_{12}|^2-n_{11}n_{22})\big]\bigg[\theta_{12}|\theta_{21}|^2-\theta_{11}\theta_{21}^*\theta_{22}-n_{12}\theta_{11}\theta_{22}^*+n_{11}\theta_{12}\theta_{22}^*+n_{12}\theta_{21}\theta_{22}^*-\theta_{12}\theta_{21}\theta_{22}^*\\	
&&\hspace{1.0cm}-n_{11}|\theta_{22}|^2+\theta_{11}|\theta_{22}|^2+(n_{11}\theta_{22}-n_{12}\theta_{21})\kappa_{12}+(n_{12}\theta_{11}-n_{11}\theta_{12})\kappa_{22}+n_{12}\big(-|\theta_{21}|^2-n_{11}\kappa_{12}+\theta_{21}\kappa_{12}\\        	
&&\hspace{1.0cm}+\theta_{11}(\theta_{21}^*-\kappa_{22})+n_{11}\kappa_{22}\big)+n_{12^*}\big(\theta_{21}^*\theta_{22}-\theta_{12}\theta_{21}^*+n_{12}\kappa_{12}-\theta_{22}\kappa_{12}-n_{12}\kappa_{22}+\theta_{12}\kappa_{22}\big)\bigg],\\
&&e^{\mu_{25}}=b^*\al_1^{(2)}|\al_1^{(1)}|^2|\al_2^{(1)}|^2|\al_2^{(2)}|^2\big[|b|^2(\theta_{12}-\theta_{22})(\theta_{12}^*\theta_{21}^*-\theta_{11}^*\theta_{22}^*)ac\big(n_{12}\theta_{22}^*(\kappa_{11}-\kappa_{12}^*)+n_{22}\theta_{21}^*(\kappa_{12}^*-\kappa_{11})\\	
&&\hspace{1.0cm}+n_{22}\theta_{11}^*(\kappa_{12}-\kappa_{22})+n_{12}\theta_{12}^*(\kappa_{22}-\kappa_{12})\big)\big]\bigg[\theta_{11}^*|\theta_{22}|^2-n_{22}\theta_{21}^*\kappa_{11}+\theta_{21}^*\theta_{22}\kappa_{11}+n_{12}\theta_{22}^*\kappa_{11}-|\theta_{22}|^2\kappa_{11}\\	
&&\hspace{1.0cm}+(n_{12}\theta_{11}^*-\theta_{11}^*\theta_{22})\kappa_{12}+(n_{22}\theta_{21}^*-n_{12}\theta_{22}^*)\kappa_{12}^*+(n_{12}-n_{22})|\kappa_{12}|^2+(-n_{22}\theta_{11}^*-n_{12}\kappa_{11}+n_{22}\kappa_{11})\kappa_{22}\\	
&&\hspace{1.0cm}+\theta_{12}^*\big(-\theta_{21}^*\theta_{22}-n_{12}\kappa_{12}+\theta_{22}\kappa_{12}+n_{12}\kappa_{22}\big)+\theta_{12}\big(-\theta_{11}^*\theta_{22}^*-\theta_{21}^*\kappa_{12}^*+\theta_{22}^*\kappa_{12}^*+\theta_{12}^*(\theta_{21}^*-\kappa_{22})+\theta_{11}^*\kappa_{22}\big)\bigg],\\
\end{eqnarray*}
\begin{eqnarray*}	
&&e^{\lam_{36}}=|\al_2^{(1)}|^2|\al_2^{(2)}|^2\big[|\theta_{22}|^2-n_{22}\kappa_{22}\big]\big[|b|^2|\theta_{22}|^2-acn_{22}\kappa_{22}\big],\\
&&e^{r_{16}}=c|\al_2^{(1)}|^2|\al_2^{(2)}|^2|\al_1^{(2)}|^2\big[n_{22}|\theta_{21}|^2-n_{12}^*\theta_{21}^*\theta_{22}-n_{12}\theta_{21}\theta_{22}^*+n_{11}|\theta_{22}|^2+|n_{12}|^2\kappa_{22}-n_{11}n_{22}\kappa_{22}\big]\\
&&\hspace{1cm}\times\big[|b|^2\big(n_{22}|\theta_{21}|^2-n_{12}^*\theta_{21}^*\theta_{22}-n_{12}\theta_{21}\theta_{22}^*+n_{11}|\theta_{22}|^2\big)+ac\kappa_{22}(|n_{12}|^2-n_{11}n_{22})\big],\\
&&e^{r_{15}}=b\al_1^{(1)}\al_1^{(2)*}|\al_2^{(1)}|^2|\al_2^{(2)}|^2 \big[\theta_{12}\theta_{21}\theta_{22}^*-\theta_{11}|\theta_{22}|^2-n_{22}\theta_{21}\kappa_{12}+n_{12}^*\theta_{22}\kappa_{12}+(n_{22}\theta_{11}-n_{12}^*\theta_{12})\kappa_{22}\big]\\	
&&\hspace{1.0cm}\times\big[|b|^2\theta_{22}^*(\theta_{12}\theta_{21}-\theta_{11}\theta_{22})+ac(n_{12}^*\theta_{22}\kappa_{12}-n_{22}\theta_{21}\kappa_{12}+(n_{22}\theta_{11}-n_{12}^*\theta_{12})\kappa_{22})\big],\\	
&&e^{r_{14}}=b^*\al_1^{(1)*}\al_1^{(2)}|\al_2^{(1)}|^2|\al_2^{(2)}|^2\big[\theta_{12}^*\theta_{21}^*\theta_{22}-\theta_{11}^*|\theta_{22}|^2-n_{22}\theta_{21}^*\kappa_{12}^*+n_{12}\theta_{22}^*\kappa_{12}^*+(n_{22}\theta_{11}^*-n_{12}\theta_{12}^*)\kappa_{22}\big]\\	
&&\hspace{1.0cm}\times\big[|b|^2\theta_{22}(\theta_{12}^*\theta_{21}^*-\theta_{11}^*\theta_{22}^*)+ac\big((n_{12}\theta_{22}^*-n_{22}\theta_{21}^*)\kappa_{12}^*+\kappa_{22}(n_{22}\theta_{11}^*-n_{12}\theta_{12}^*)\big)\big],\\
&&e^{r_{13}}=a|\al_1^{(1)}|^2|\al_2^{(1)}|^2|\al_2^{(2)}|^2\big[|\theta_{22}|^2\kappa_{11}-\theta_{12}^*\theta_{22}\kappa_{12}-\theta_{12}\theta_{22}^*\kappa_{12}^*+n_{22}|\kappa_{12}|^2+(|\theta_{12}|^2-n_{22}\kappa_{11})\kappa_{22}\big]\\	
&&\hspace{1cm}\times\big[|b|^2(|\theta_{22}|^2\kappa_{11}-\theta_{12}^*\theta_{22}\kappa_{12}-\theta_{12}\theta_{22}^*\kappa_{12}^*+|\theta_{12}|^2\kappa_{22})+acn_{22}(|\kappa_{12}|^2-\kappa_{11}\kappa_{22})\big],\\
&&e^{\chi_{23}}=ab\al_1^{(1)}|\al_2^{(1)}|^2|\al_2^{(2)}|^2(\theta_{22}\kappa_{12}-\theta_{12}\kappa_{22})\big[\theta_{22}(\kappa_{12}-\theta_{22}^*)+\theta_{12}(\theta_{22}^*-\kappa_{22})+n_{22}(\kappa_{22}-\kappa_{12})\big],\\
&&e^{\chi_{24}}=-\al_1^{(2)}|\al_2^{(1)}|^2|\al_2^{(2)}|^2\big[|b|^2\theta_{22}(\theta_{21}^*-\theta_{22}^*)+ac\kappa_{22}(n_{22}-n_{12})\big]\\	
&&\hspace{2.0cm}\times\big[-n_{12}\theta_{22}^*-\theta_{21}^*\theta_{22}+|\theta_{22}|^2+n_{22}(\theta_{21}^*-\kappa_{22})+n_{12}\kappa_{22}\big],\\
%\end{eqnarray*}
%\begin{eqnarray*}	
&&e^{\chi_{25}}=b\al_1^{(1)}|\al_1^{(2)}|^2|\al_2^{(1)}|^2|\al_2^{(2)}|^2\big[|b|^2(\theta_{12}\theta_{21}-\theta_{11}\theta_{22})(\theta_{21}^*-\theta_{22}^*)-ac\big((n_{12}-n_{22})\theta_{21}\kappa_{12}+\theta_{22}\kappa_{12}(n_{12}^*-n_{11})\\	
&&\hspace{1.0cm}+\kappa_{22}\theta_{11}(n_{22}-n_{12}\theta_{12})+\theta_{12}\kappa_{22}(n_{11}-n_{12}^*)\big)\big]\bigg[\theta_{12}|\theta_{21}|^2-\theta_{11}\theta_{21}^*\theta_{22}-n_{12}\theta_{11}\theta_{22}^*+\theta_{22}^*(n_{11}\theta_{12}+n_{12}\theta_{21}\\	
&&\hspace{1.0cm}-\theta_{12}\theta_{21})+(\theta_{11}-n_{11})|\theta_{22}|^2+(n_{11}\theta_{22}-n_{12}\theta_{21})\kappa_{12}+(n_{12}\theta_{11}-n_{11}\theta_{12})\kappa_{22}+n_{22}\big(-|\theta_{21}|^2+(\theta_{21}-n_{11})\kappa_{12}\\
&&\hspace{1cm}+n_{11}\kappa_{22}+\theta_{11}(\theta_{21}^*-\kappa_{22})\big)+n_{12}^*\big(-\theta_{12}\theta_{21}^*+\theta_{21}^*\theta_{22}+(n_{12}-\theta_{22})\kappa_{12}+(\theta_{12}-n_{12})\kappa_{22}\big)\bigg],\\
&&e^{\chi_{28}}=-a|\al_1^{(1)}|^2|\al_2^{(1)}|^2|\al_2^{(2)}|^2\al_1^{(2)}\big[|b|^2\big(\kappa_{11}(|\theta_{22}|^2-\theta_{21}\theta_{22})+\kappa_{12}(\theta_{11}^*\theta_{22}-\theta_{12}^*\theta_{22})+(\theta_{12}\theta_{21}^*-\theta_{12}\theta_{22}^*)\kappa_{12}^*\\
&&\hspace{1cm}+(|\theta_{12}|^2-\theta_{11}^*\theta_{12})\kappa_{22}\big)-ac(n_{12}-n_{22})(|\kappa_{12}|^2-\kappa_{11}\kappa_{22})\big]\bigg[\theta_{11}^*|\theta_{22}|^2+\kappa_{11}(\theta_{21}^*\theta_{22}-|\theta_{22}|^2-n_{22}\theta_{21}^*+n_{12}\theta_{22}^*)\\
&&\hspace{1.0cm}+\kappa_{12}\theta_{11}^*(n_{22}-\theta_{22})+\kappa_{12}^*(n_{22}\theta_{21}^*-n_{12}\theta_{22}^*)+|\kappa_{12}|^2(n_{12}-n_{22})-\kappa_{22}(n_{22}\theta_{11}^*+n_{12}\kappa_{11}-n_{22}\kappa_{11})\\	
&&\hspace{1.0cm}+\theta_{12}^*(-\theta_{21}^*\theta_{22}-n_{12}\kappa_{12}+\theta_{22}\kappa_{12}+n_{12}\kappa_{22})+\theta_{12}(-\theta_{11}^*\theta_{22}^*+(\theta_{22}^*-\theta_{21}^*)\kappa_{12}^*+\theta_{12}^*(\theta_{21}^*-\kappa_{22})+\theta_{11}^*\kappa_{22})\bigg],\\
&&e^{\ga_{15}}=b^*|\al_2^{(1)}|^2|\al_2^{(2)}|^2(\theta_{22}^*-\kappa_{22}),~e^{\ga_{20}}=cn_{22}\al_2^{(1)}|\al_2^{(2)}|^2(n_{22}-\theta_{22}),~e^{\del_4}=a|\al_2^{(1)}|^2\kappa_{22},~e^{\del_{11}}=b\al_2^{(1)}\al_2^{(2)*}\theta_{22}^2,\\	
&&e^{\del_{16}}=cn_{22}^2|\al_2^{(2)}|^2,~e^{\del_{12}}=b^*\al_2^{(1)*}\al_2^{(2)}\theta_{22}^{*2},~e^{\nu_{20}}=-b\al_2^{(1)}|\al_2^{(2)}|^2\theta_{22}(n_{22}-\theta_{22}),~e^{\nu_{15}}=-a|\al_2^{(1)}|^2\al_2^{(2)}\kappa_{22}(\theta_{22}^*-\kappa_{22}).
\end{eqnarray*}

\section*{Appendix C: The various constants which appear in Section 4}\label{C}
\begin{eqnarray*}
&&e^{\hat{\ga_1}}=e^{\ga_3}+e^{\ga_4}+e^{\ga_5}+e^{\ga_6},~	e^{\hat{\ga_2}}=e^{\ga_{10}}+e^{\ga_{11}},~e^{\hat{\mu_2}}=e^{\mu_9}+e^{\mu_{10}}+e^{\mu_{11}}+e^{\mu_{12}},~e^{\Delta_{11}}=e^{\mu_{23}}+e^{\mu_{24}},\\&&	e^{\hat{\mu_1}}=e^{\mu_{18}}+e^{\mu_{19}}+e^{\mu_{20}}+e^{\mu_{21}},~e^{\Delta_{12}}=e^{\chi_{23}}+e^{\chi_{24}},~e^{\hat{\lam}_1}=e^{\lam_6}+e^{\lam_7}+e^{\lam_8}+e^{\lam_9},~e^{\hat{\nu}_1}=e^{\nu_3}+e^{\nu_4},\\
&&e^{\hat{\lam}_2}=e^{\lam_{21}}+e^{\lam_{22}}+e^{\lam_{23}}+e^{\lam_{24}},~	
e^{\hat{\lam}_3}=e^{\lam_{13}}+e^{\lam_{14}}+e^{\lam_{15}}+e^{\lam_{16}}~
e^{\hat{\lam}_4}=e^{\lam_{28}}+e^{\lam_{29}}+e^{\lam_{30}}+e^{\lam_{31}}\\	
&&e^{\hat{\lam}_5}=e^{r_{13}}+e^{r_{14}}+e^{r_{15}}+e^{r_{16}},~
e^{\hat{\nu}_2}=e^{\nu_7}+e^{\nu_8}+e^{\nu_9}+e^{\nu_{10}},~
e^{\hat{\chi}_2}=e^{\chi_9}+e^{\chi_{10}}+e^{\chi_{11}}+e^{\chi_{12}},\\
&&e^{\hat{\chi}_1}=e^{\chi_{18}}+e^{\chi_{19}}+e^{\chi_{20}}+e^{\chi_{21}},\\
\end{eqnarray*}
\begin{eqnarray*}
&&e^{\ga_3}=\frac{a(k_1-k_2)^2\al_2^{(1)}|\al_1^{(1)}|^2}{(k_1+k_1^*)^2(k_2+k_1^*)^2},~
e^{\ga_4}=\frac{b^*(k_2-k_1)^2\al_1^{(2)}\al_2^{(1)}\al_1^{(1)*}}{(k_1+k_1^*)^2(k_2+k_1^*)},~
e^{\ga_5}=\frac{b(k_1-k_2)^2\al_2^{(1)}\al_1^{(1)}\al_1^{(2)*}}{(k_1+k_1^*)^2(k_2+k_1^*)^2},\\
&&e^{\ga_6}=\frac{c(k_2-k_1)|\al_1^{(2)}|^2\al_2^{(1)}}{(k_1+k_1^*)^2(k_2+k_1^*)},~e^{\ga_{10}}=\frac{b^*(k_1-l_2)|\al_1^{(1)}|^2\al_2^{(2)}}{(k_1+k_1^*)(l_2+k_1^*)^2},~e^{\ga_{11}}=\frac{c(k_1-l_2)\al_1^{(1)}\al_1^{(2)*}\al_2^{(2)}}{(k_1+k_1^*)(l_2+k_1^*)^2},\\
&&e^{\nu_3}=\frac{a(k_1-k_2)\al_1^{(1)*}\al_1^{(2)}\al_2^{(1)}}{(k_2+k_1^*)^2(k_1+k_1^*)},~e^{\nu_4}=\frac{b(k_1-k_2)|\al_1^{(2)}|^2\al_2^{(1)}}{(k_2+k_1^*)^2(k_1+k_1^*)},~e^{\nu_7}=-\frac{a(k_1-l_2)|\al_1^{(1)}|^2\al_2^{(2)}}{(l_2+k_1^*)(k_1+k_1^*)^2},\\
&&e^{\nu_8}=\frac{b^*(k_1-l_2)^2\al_1^{(1)*}\al_1^{(2)}\al_2^{(2)}}{(l_2+k_1^*)^2(k_1+k_1^*)^2},~e^{\nu_9}=-\frac{b^*(k_1-l_2)^2\al_1^{(1)*}\al_1^{(2)}\al_2^{(2)}}{(l_2+k_1^*)^2(k_1+k_1^*)^2},~e^{\nu_{10}}=\frac{c(k_1-l_2)^2|\al_1^{(2)}|^2\al_2^{(2)}}{(l_2+k_1^*)^2(k_1+k_1^*)^2},\\
&&e^{\mu_9}=\frac{ab^*|k_1-k_2|^4(k_1-l_2)(k_2-l_2)[k_1(k_2-l_2)-l_2(k_2+k_2^*)-k_1^*(k_2^*+l_2)]|\al_1^{(1)}|^2|\al_2^{(1)}|^2\al_2^{(2)}}{(k_1+k_1^*)^2|k_1+k_2^*|^4(k_2+k_2^*)^2(k_1^*+l_2)^2(k_2^*+l_2)^2},\\
&&e^{\mu_{10}}=\frac{b^{*2}(k_1^*-k_2^*)^2(k_2-k_1)(k_2-l_2)(k_1-l_2)^2\al_1^{(1)*}\al_1^{(2)}\al_2^{(2)}|\al_2^{(1)}|^2}{(k_1^*+k_2)(k_2+k_2^*)(k_1+k_1^*)^2(k_1+k_2^*)^2(k_1^*+l_2)^2(k_2^*+l_2)^2},\\
%\end{eqnarray*}
%\begin{eqnarray*}
&&e^{\mu_{11}}=\frac{(k_1-k_2)^2(k_2^*-k_1^*)(k_1-l_2)(k_2-l_2)\Lam_1\al_1^{(1)}\al_1^{(2)*}|\al_2^{(1)}|^2\al_2^{(2)}}{(k_1+k_1^*)^2(k_1+k_2^*)^2(k_2+k_2^*)^2(k_2+k_1^*)^2(k_2^*+l_2)^2(k_1^*+l_2)^2},\\
&&\Lam_1=\big[ac(k_1+k_1^*)(k_2+k_1^*)(k_2^*+l_2)-|b|^2(k_1+k_2^*)(k_2+k_2^*)(k_1^*+l_2)\big],\\
&&e^{\mu_{12}}=\frac{b^*c(k_2-k_1)(k_2^*-k_1^*)^2(k_2-l_2)(k_1-l_2)^2|\al_1^{(2)}|^2|\al_2^{(1)}|^2\al_2^{(2)}}{(k_2+k_2^*)(k_2^*+k_1)^2(k_2+k_1^*)(k_1+k_1^*)^2(k_2^*+l_2)^2(k_1^*+l_2)^2},\\
&&e^{\mu_{18}}=\frac{(k_1-k_2)^2|k_1-l_2|^2(k_2-l_2)\Lam_2|\al_1^{(1)}|^2|\al_2^{(2)}|^2\al_2^{(1)}}{(k_1+k_1^*)^2(k_1^*+k_2)^2|k_1+l_2^*|^4(k_2+l_2^*)^2(l_2+l_2^*)^2},~e^{\mu_{19}}=\frac{b^*c(k_2-k_1)(k_2-l_2)|k_1-l_2|^4\al_1^{(1)*}\al_1^{(2)}\al_2^{(1)}|\al_2^{(2)}|^2}{(k_1^*+k_2)(k_1+k_1^*)^2|k_1+l_2^*|^4(k_2+l_2^*)(l_2+l_2^*)^2},\\
&&e^{\mu_{21}}=\frac{c^2(k_2-k_1)(k_2-l_2)|k_1-l_2|^4|\al_1^{(2)}|^2\al_2^{(1)}|\al_2^{(2)}|^2}{(k_2+k_1^*)(k_1+k_1^*)^2|k_1+l_2^*|^4(k_2+l_2^*)(l_2+l_2^*)^2},~e^{\mu_{23}}=\frac{(k_1-k_2)^2(k_1-l_2)|k_2-l_2|^2\Lam_3\al_1^{(1)}|\al_2^{(1)}|^2|\al_2^{(2)}|^2}{(k_1+k_2^*)^2(k_2+k_2^*)^2|k_2+l_2^*|^4(k_1+l_2^*)^2(l_2+l_2^*)^2},\\
&&e^{\mu_{24}}=\frac{b^*c(k_2-k_1)(k_2-l_2)(k_1-l_2)^2(k_2^*-l_2^*)^2\al_1^{(2)}|\al_2^{(1)}|^2|\al_2^{(2)}|^2}{(k_2+k_2^*)(k_2^*+k_1)^2(k_2^*+l_2)^2(k_2+l_2^*)(k_1+l_2^*)^2(l_2+l_2^*)^2},\\
&&e^{\mu_{20}}=-\frac{bc(k_1-k_2)^2(k_1-l_2)(k_2-l_2)(k_1^*-l_2^*)^2[l_2(k_2+k_1^*)+k_1(l_2-k_2)+l_2^*(k_1^*+l_2)]\al_1^{(1)}\al_1^{(2)*}\al_2^{(1)}|\al_2^{(2)}|^2}{(k_1+k_1^*)^2(k_2+k_1^*)^2|k_1+l_2^*|^4(k_2+l_2^*)^2(l_2+l_2^*)^2},\\
&&e^{\chi_9}=\frac{a^2|k_1-k_2|^4(k_1-l_2)(k_2-l_2)|\al_1^{(1)}|^2|\al_2^{(1)}|^2\al_2^{(2)}}{(k_1+k_1^*)^2|k_1+k_2^*|^4(k_2+k_2^*)^2(k_1^*+l_2)(k_2^*+l_2)},\\
&&e^{\chi_{11}}=\frac{ab|k_1-k_2|^4(k_1-l_2)(k_2-l_2)\al_1^{(1)}\al_1^{(2)*}|\al_2^{(1)}|^2\al_2^{(2)}}{(k_1+k_2^*)^2(k_2+k_2^*)^2(k_1+k_1^*)^2(k_2+k_1^*)^2(k_2^*+l_2)(k_1^*+l_2)},\\
&&e^{\chi_{10}}=\frac{ab^*(k_1^*-k_2^*)^2(k_2-k_1)(k_2-l_2)(k_1-l_2)^2[k_1l_2-k_1^*(k_2+k_2^*)-k_2(k_2^*+k_1+l_2)]|\al_2^{(1)}|^2\al_1^{(1)*}\al_1^{(2)}\al_2^{(2)}}{(k_1^*+k_2)^2(k_2+k_2^*)^2(k_1+k_1^*)^2(k_1+k_2^*)^2(k_1^*+l_2)^2(k_2^*+l_2)^2},\\
&&e^{\chi_{18}}=\frac{ab(k_1-k_2)^2(k_1-l_2)(k_2-l_2)(k_1^*-l_2^*)^2|\al_1^{(1)}|^2|\al_2^{(2)}|^2\al_2^{(1)}}{(k_1+k_1^*)^2(k_1^*+k_2)^2(k_1^*+l_2)(k_1+l_2^*)^2(l_2+l_2^*)(k_2+l_2^*)^2},~e^{\lam_6}=\frac{a^2|k_1-k_2|^4|\al_1^{(1)}|^2|\al_2^{(1)}|^2}{(k_1+k_1^*)^2|k_1+k_2^*|^4(k_2+k_2^*)^2},\\
&&e^{\chi_{20}}=\frac{b^2(k_1-k_2)^2(k_1-l_2)(k_2-l_2)(k_1^*-l_2^*)^2\al_1^{(1)}\al_1^{(2)*}\al_2^{(1)}|\al_2^{(2)}|^2}{(k_1+k_1^*)^2(k_2+k_1^*)^2(k_1^*+l_2)(k_1+l_2^*)^2(k_2+l_2^*)^2(l_2+l_2^*)},\\
&&e^{\chi_{12}}=\frac{|k_1-k_2|^2(k_2-l_2)(k_1-l_2)^2\Lam_4\al_1^{(1)*}\al_1^{(2)}\al_2^{(1)}|\al_2^{(2)}|^2}{(k_2^*+k_2)^2(l_2+k_2^*)^2|k_1+k_2^*|^4(k_1+k_1^*)^2(k_1^*+l_2)^2},\\
&&\Lam_4=[ac|k_1+k_2^*|^2(k_2^*+l_2)-|b|^2(k_2+k_2^*)(k_1+k_1^*)(k_1^*+l_2)],\\
&&e^{\chi_{19}}=\frac{(k_1-k_2)(k_2-l_2)(k_1-l_2)^2(k_1^*-l_2^*)\Lam_5\al_1^{(1)*}\al_1^{(2)}\al_2^{(1)}|\al_2^{(2)}|^2}{(k_1^*+k_2)^2(k_1+k_1^*)^2|k_1+l_2^*|^4(k_2+l_2^*)^2(l_2+l_2^*)^2},~
e^{\lam_7}=\frac{ab^*|k_1-k_2|^4\al_1^{(1)*}\al_1^{(2)}|\al_2^{(1)}|^2}{(k_1^*+k_2)^2(k_2+k_2^*)^2(k_1+k_1^*)^2(k_1+k_2^*)^2},\\
&&\Lam_5=[ac(k_1+k_1^*)(k_1^*+l_2)(k_2+k_2^*)-|b|^2(k_1^*+k_2)(k_1+l_2^*)(l_2+l_2^*)],\\
%\end{eqnarray*}
%\begin{eqnarray*}
&&e^{\lam_8}=\frac{ab|k_1-k_2|^4\al_1^{(1)}\al_1^{(2)*}|\al_2^{(1)}|^2}{(k_1+k_2^*)^2(k_2+k_2^*)^2(k_1+k_1^*)^2(k_2+k_1^*)^2},\\
&&e^{\chi_{23}}=\frac{ab(k_1-k_2)^2(k_1-l_2)(k_2-l_2)(k_2^*-l_2^*)^2\al_1^{(1)}|\al_2^{(1)}|^2|\al_2^{(2)}|^2}{(k_1+k_2^*)^2(k_2+k_2^*)^2(k_2^*+l_2)(k_1+l_2^*)^2(k_2+l_2^*)^2(l_2+l_2^*)},~e^{\lam_{13}}=\frac{ab^*(k_1^*-k_2^*)^2(k_1-l_2)^2|\al_1^{(1)}|^2\al_2^{(1)*}\al_2^{(2)}}{(k_1+k_1^*)^2(k_1+k_2^*)^2(k_1^*+l_2)^2(k_2^*+l_2)^2},\\
\end{eqnarray*}
\begin{eqnarray*}
&&e^{\chi_{21}}=\frac{bc(k_1-k_2)(k_2-l_2)|k_1-l_2|^4|\al_1^{(2)}|^2|\al_2^{(2)}|^2\al_2^{(1)}[k_2(k_1+k_1^*+l_2+l_2^*)-k_1l_2+k_1^*l_2^*]}{(k_2+k_1^*)^2(k_1+k_1^*)^2|k_1+l_2^*|^4(k_2+l_2^*)^2(l_2+l_2^*)^2},\\
&&e^{\chi_{24}}=\frac{(k_1-k_2)|k_2-l_2|^2(k_1-l_2)^2\Lam_6\al_1^{(2)}|\al_2^{(1)}|^2|\al_2^{(2)}|^2}{(k_2+k_2^*)^2(k_1+k_2^*)^2|k_2+l_2^*|^4(k_1+l_2^*)^2(l_2+l_2^*)^2},\\
&&\Lam_6=[ac(k_1+k_2^*)|k_2+l_2^*|^2-|b|^2(k_2+k_2^*)(k_1+l_2^*)(l_2+l_2^*)],\\
&&e^{\lam_9}=\frac{|k_1-k_2|^2[ac|k_1+k_2^*|^2-|b|^2(k_1+k_1^*)(k_2+k_2^*)]|\al_1^{(2)}|^2|\al_2^{(1)}|^2}{(k_2+k_2^*)^2(k_1+k_1^*)^2|k_1+k_2^*|^4},\\
&&e^{\lam_{14}}=\frac{b^{*2}(k_1^*-k_2^*)^2(k_1-l_2)^2\al_1^{(1)*}\al_1^{(2)}\al_2^{(1)*}\al_2^{(2)}}{(k_1+k_1^*)^2(k_1+k_2^*)^2(k_1^*+l_2)^2(k_2^*+l_2)^2},\\
&&e^{\lam_{15}}=\frac{(k_2^*-k_1^*)(k_1-l_2)\Lam_7\al_1^{(1)}\al_1^{(2)*}\al_2^{(1)*}\al_2^{(2)}}{(k_1+k_2^*)^2(k_1+k_1^*)^2(k_2^*+l_2)^2(k_1^*+l_2)^2},~\Lam_7=[ac(k_1+k_1^*)(k_2^*+l_2)-|b|^2(k_1+k_2^*)(k_1^*+l_2)],\\
&&e^{\lam_{16}}=\frac{b^*c(k_1^*-k_2^*)^2(k_1-l_2)^2|\al_1^{(2)}|^2\al_2^{(1)*}\al_2^{(2)}}{(k_1+k_2^*)^2(k_1+k_1^*)^2(k_2^*+l_2)^2(k_1^*+l_2)^2},~e^{\lam_{21}}=\frac{ab(k_1-k_2)^2(k_1^*-l_2^*)^2|\al_1^{(1)}|^2\al_2^{(1)}\al_2^{(2)*}}{(k_1+k_1^*)^2(k_1^*+k_2)^2(k_1+l_2^*)^2(k_2+l_2^*)^2},\\
&&e^{\lam_{23}}=\frac{b^2(k_1-k_2)^2(k_1^*-l_2^*)^2\al_1^{(1)}\al_1^{(2)*}\al_2^{(1)}\al_2^{(2)*}}{(k_1+k_1^*)^2(k_2+k_1^*)^2(k_1+l_2^*)^2(k_2+l_2^*)^2},~e^{\lam_{24}}=\frac{bc(k_1-k_2)^2(k_1^*-l_2^*)^2|\al_1^{(2)}|^2\al_2^{(1)}\al_2^{(2)*}}{(k_2+k_1^*)^2(k_1+k_1^*)^2(k_2+l_2^*)^2(k_1+l_2^*)^2},\\
&&e^{\lam_{22}}=\frac{(k_2-k_1)(k_1^*-l_2^*)[ac(k_1+k_1^*)(k_2+l_2^*)-|b|^2(k_1^*+k_2)(k_1+l_2^*)]\al_1^{(1)*}\al_1^{(2)}\al_2^{(1)}\al_2^{(2)*}}{(k_1^*+k_2)^2(k_1+k_1^*)^2(k_2+l_2^*)^2(k_1+l_2^*)^2},\\
&&e^{\lam_{28}}=\frac{|k_1-l_2|^2[ac|k_1+l_2^*|^2-|b|^2(k_1+k_1^*)(l_2+l_2^*)]|\al_1^{(1)}|^2|\al_2^{(2)}|^2}{(k_1+k_1^*)^2(l_2+l_2^*)^2|k_1+l_2^*|^4},~e^{\lam_{29}}=\frac{b^*c|k_1-l_2|^4\al_1^{(1)*}\al_1^{(2)}|\al_2^{(2)}|^2}{(k_1+k_1^*)^2(k_1^*+l_2)^2(k_1+l_2^*)^2(l_2+l_2^*)^2},\\
&&e^{\lam_{30}}=\frac{bc|k_1-l_2|^4\al_1^{(1)}\al_1^{(2)*}|\al_2^{(2)}|^2}{(k_1+k_1^*)^2|k_1+l_2^*|^4(l_2+l_2^*)^2},~e^{\lam_{31}}=\frac{c^2|k_1-l_2|^4|\al_1^{(2)}|^2|\al_2^{(2)}|^2}{(k_1+k_1^*)^2|k_1+l_2^*|^4(l_2+l_2^*)^2},\\
&&e^{r_{13}}=\frac{a|k_1-k_2|^4|k_1-l_2|^2|k_2-l_2|^2|\al_1^{(1)}|^2|\al_2^{(1)}|^2|\al_2^{(2)}|^2\Lam_8}{(k_1+k_1^*)^2|k_1+k_2^*|^4(k_2+k_2^*)^2|k_1+l_2^*|^4|k_2+l_2^*|^4(l_2+l_2^*)^2},\\
&&e^{r_{14}}=\frac{b^*(k_1^*-k_2^*)^2(k_2-k_1)|k_2-l_2|^2(k_1-l_2)^2(k_1^*-l_2^*)|\al_2^{(1)}|^2|\al_2^{(2)}|^2\al_1^{(1)*}\al_1^{(2)}\Lam_9}{(k_1^*+k_2)^2(k_2+k_2^*)^2(k_1+k_1^*)^2(k_1+k_2^*)^2|k_1+l_2^*|^4|k_2+l_2^*|^4(l_2+l_2^*)^2},\\
&&e^{r_{15}}=\frac{b(k_1-k_2)^2(k_2^*-k_1^*)(k_1-l_2)(k_1^*-l_2^*)^2|k_2-l_2|^2\al_1^{(1)}\al_1^{(2)*}|\al_2^{(1)}|^2|\al_2^{(2)}|^2\Lam_{10}}{(k_1+k_2^*)^2(k_2+k_2^*)^2(k_1+k_1^*)^2(k_2+k_1)^2|k_2+l_2^*|^4|k_1+l_2^*|^4(l_2+l_2^*)^2},\\
&&e^{r_{16}}=\frac{c|k_1-k_2|^2|k_2-l_2|^2|k_1-l_2|^4|\al_1^{(2)}|^2|\al_2^{(1)}|^2|\al_2^{(2)}|^2\Lam_{11}}{(k_2+k_2^*)^2|k_1+k_2^*|^4(k_1+k_1^*)^2|k_2+l_2^*|^4|k_1+l_2^*|^4(l_2+l_2^*)^2},\\
&&\Lam_8=\big[ac|k_1+l_2^*|^2|k_2+l_2^*|^2-|b|^2(l_2+l_2^*)\big(l_2l_2^*(k_2+k_2^*)+k_1k_1^*(k_2+k_2^*)+k_1k_2(k_2^*+l_2-l_2^*)+l_2l_2^*\\
&&\hspace{0.5cm}+k_1^*(k_2k_2^*+l_2l_2^*+k_2^*(l_2^*-l_2))\big)\big],\\
&&\Lam_9=\big[|b|^2(k_1^*+k_2)(k_2+k_2^*)(k_1+l_2^*)(l_2+l_2^*)-ac(k_2+l_2^*)\big(k_1l_2(k_2^*-l_2^*)+k_2k_1(l_2+l_2^*)\\
&&\hspace{0.5cm}+k_1^*(k_1l_2+k_2(l_2^*-k_2^*)+k_2^*(k_1+l_2+l_2^*))\big)\big],\\
%\end{eqnarray*}
%\begin{eqnarray*}
&&\Lam_{10}=\big[|b|^2(k_1+k_2^*)(k_2+k_2^*)(k_1^*+l_2)(l_2+l_2^*)-ac(k_2^*+l_2)\big(l_2k_2^*(k_2+k_1^*)+k_1^*l_2^*(k_2+k_2^*)\\
&&\hspace{0.5cm}+l_2l_2^*(k_2^*-k_1^*)+k_1(k_2^*l_2+k_1^*l_2^*+k_2(l_2+l_2^*+k_1^*-k_2^*))\big)\big],\\
&&\Lam_{11}=\big[ac|k_1+k_2^*|^2|k_2+l_2^*|^2-|b|^2(k_2+k_2^*)\big(l_2k_1(k_1^*-k_2^*)+l_2^*k_1^*(k_1+k_2^*)+l_2l_2^*(k_1+k_1^*)\\
&&\hspace{0.5cm}+k_1(k_1l_2-k_1^*l_2^*+k_2^*(k_1+k_1^*+l_2+l_2^*))\big)\big].
\end{eqnarray*}




\end{document}